\documentstyle[prd,aps,preprint,eqsecnum,epsfig,floats]{revtex}
\def\bege{\begin{equation}}
\def\ende{\end{equation}}

\def\bar#1{\overline{ #1 }}

\def\beq{\begin{equation}}
\def\eeq{\end{equation}}
\def\bea{\begin{eqnarray}}
\def\eea{\end{eqnarray}}

\draft

\begin{document}
\tighten

\title{Regularizing the Divergent Structure of the Light-Front Currents}

\author{Bernard L. G. Bakker$^1$, Ho-Meoyng Choi$^{2,3}$,
and Chueng-Ryong Ji$^2$}
\address{$^1$ Department of Physics and Astrophysics, Vrije Universiteit,
De Boelelaan 1081, NL-1081 HV Amsterdam, The Netherlands}
\address{$^2$ Department of Physics, North Carolina State University,
Raleigh, NC 27695-8202, USA}
\address{$^3$ Physics Department, Carnegie Mellon University, 
Pittsburgh, PA 15213, USA}


\maketitle

\begin{abstract}

The divergences appearing in the 3+1 dimensional fermion-loop calculations
are often regulated by smearing the vertices in a covariant manner. 
Performing a parallel light-front calculation, we corroborate the similarity
between the vertex-smearing-technique and the Pauli-Villars regularization.
In the light-front calculation of the electromagnetic meson current, 
we find that the persistent end-point singularity that appears in the 
case of point vertices is removed
even if the smeared-vertex is taken to the limit of the point-vertex.
Recapitulating the current conservation, we substantiate the finiteness
of both valence and non-valence contributions in all components of the current 
with the regularized bound-state vertex. 
However, we stress that each contribution, valence or non-valence,
depends on the reference-frame even though the sum is always frame-independent.
The numerical taxonomy of each contribution including
the instantaneous contribution and the zero-mode contribution
is presented in the $\pi$, $K$, and $D$-meson form factors.  

\end{abstract}
\pacs{ }

\section{Introduction}
\label{sec.1}

With the recent advances in the Hamiltonian renormalization program,
Light-Front Dynamics (LFD) appears to be a promising technique 
to impose the relativistic treatment of hadrons.
In LFD a Fock-space expansion of bound states is made to handle 
the relativistic many-body effects in a consistent way \cite{BPP}.
The wave function $\psi_n(x_i, \vec{k}_{i\,\perp}, \lambda_i)$ describes the
component with $n$ constituents, with longitudinal momentum fraction
$x_i$, perpendicular momentum $\vec{k}_{i\,\perp}$ and helicity $\lambda_i$,
$i=1, \dots, n$. It is the aim of LFD to determine those wave functions
and use them in conjunction with hard scattering amplitudes to describe
the properties of hadrons and their response to electroweak probes.
Important steps were taken towards a realization of
this goal\cite{Hil}.
However, at present
there are no realistic results available for wave functions of hadrons
based on QCD alone. In order to calculate the response of hadrons to external
probes, one might resort to the use of model wave functions.
The variational principle enabled the solution of a QCD-motivated effective
Hamiltonian and the constructed LF quark-model provided a good description of 
the available experimental data spanning various meson properties \cite{Ji}.
The same reasons that make LFD so attractive to solve
bound-state problems in field theory make it also useful for a
relativistic description of nuclear systems. Presently, it is realized
that a parametrization of nuclear reactions in terms of
non-relativistic wave functions must fail.  LF methods have the
advantage that they are formally similar to time-ordered many-body
theories, yet provide relativistically invariant observables.
Furthermore, as far as the concerned amplitude is unconditionally 
(or absolutely) convergent, the LF Hamiltonian approach must yield
the same result as the one obtained by a covariant Feynman approach.

\begin{figure}
\begin{center}
\epsfig{figure=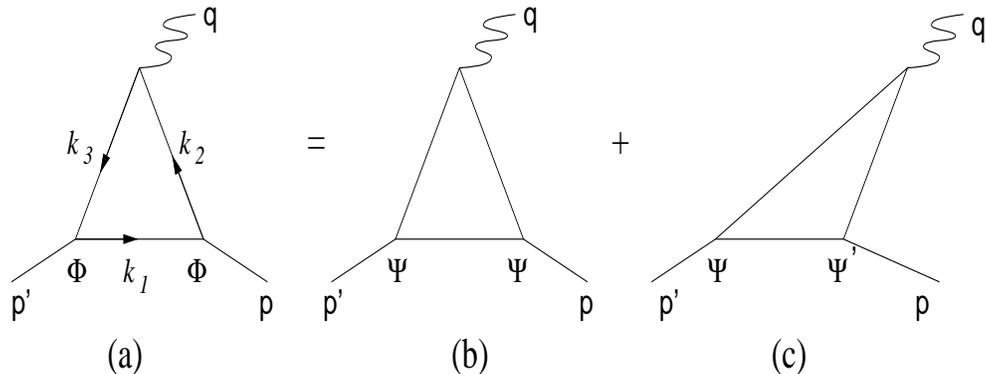,height=50mm,width=130mm}
\caption{Covariant triangle diagram(a) is equal to the sum of two 
light-front time-ordered diagrams, i.e., valence(b) and nonvalence(c) 
diagrams.\label{fig0.1}}
\end{center}
\end{figure}

However, not all is well.
As we have recently shown\cite{BJ}, the amplitudes that are only
conditionally convergent must be treated with care.
A case in point discussed in our work\cite{BJ} was the calculation
of a current matrix element in quantum field theory.
A typical amplitude is given by the triangle diagram. One encounters this
diagram e.g.  when computing the meson form factor 
(see Fig.~\ref{fig0.1}(a)).
The vertices denoted by $\Phi$ are coupling constants in covariant
perturbation theory. The hard scattering process is the absorption of a
photon of momentum $q$ by a (anti-)quark. In the LFD approach the
covariant amplitude is replaced by a series of LF time-ordered
diagrams. In the case of the triangle diagram they are depicted in
Figs.~\ref{fig0.1}(b) and~1(c).
The first (Fig.~\ref{fig0.1}(b)) of these two diagrams is easily 
interpreted in terms of the LF wave functions $\Psi$. 
However, the other diagram (Fig.~\ref{fig0.1}(c)) has one vertex that
can again be written in the same way as before, but it contains also
another vertex, denoted by $\Psi'$, that cannot be written as a LF
wave function. 
The necessity of this new element $\Psi'$ in LFD has also been discussed
in the cases of semileptonic meson decays\cite{semilep} and deeply virtual
Compton scattering\cite{compton}.
One may call $\Psi$ and $\Psi'$ the vertices of leading order and 
non-leading order, respectively, in the sense that the leading order vertex 
corresponds to the lowest Fock-state whereas the non-leading order vertex
takes into account the higher Fock-states.
The diagrams with $\Psi'$ are designated as {\em nonvalence}
diagrams while those with vertices of type $\Psi$ only are designated
as {\em valence} diagrams.
In order to obtain the invariant form factor, the two LF form factors must
be added. It depends on the situation whether one can limit oneself
to a single component of the current $J^\mu$ to extract the invariant
objects. For the electromagnetic current of a spin-0 particle any
single component would suffice to extract the unique form factor. On
the other hand, in situations like semileptonic pseudoscalar meson decay, 
which involves two independent form factors, or the electromagnetic
current of a (axial-)vector particle that is described by three
independent form factors, one must use information from several
current components to determine the invariant amplitudes.

Earlier, we presented an analysis of contributions from the
nonvalence diagrams\cite{BJ}.
We constructed both leading and non-leading order vertices using
pointlike covariant ones. 
The model that we used was essentially an extension of Mankiewicz
and Sawicki's $(1+1)$-dimensional quantum field theory model \cite{SM}, 
which was later reinvestigated by several others \cite{SB,GS,Sawicki,CJ,BH}.
While their model\cite{SM} was a simple 1+1 dim. scalar field theory,
it included a binding effect of the two-body bound-state. Indeed, in 
Ref.~\cite{GS}, the relativistic two-body bound-state form factor was 
discussed in the full range of the binding energy.
The starting model wave function was the solution of the covariant
Bethe-Salpeter equation in the ladder approximation with a
relativistic version of the contact interaction \cite{GS}.
The covariant model wave function was a product of two free single 
particle propagators, the overall
momentum-conserving Dirac delta function, and a constant vertex function.
Consequently, all our form factor calculations were 
various ways of evaluating the Feynman triangle
diagram in quantum field theory.
As pointed out in Ref.~\cite{FM}, however, the elastic electromagnetic 
form factors of a bound-state computed from the triangle diagram and from 
the Hamiltonian front-form dynamics are the same\cite{FS}.
Since our aim was to analyze the taxonomy of the triangle diagram, 
we didn't choose any particular gauge for the electromagnetic gauge field
but presented the equivalence of the physical form factor $F(q^2)$ in any 
choice of the electromagnetic gauge
\footnote{For a recent advocacy of the anti-light-cone gauge, $A^- = 0$, see
M. Morara, R. Soldati, and G. McCartor, AIP Conference Proceedings ${\bf 494}$,
284-290 (1999).}
or in any choice of the current component.

Our conditionally convergent example of LFD with a fermion-loop showed that 
the bad component of the current, $J^-$, with 
spin-1/2 constituents
exhibits a persistent end-point singularity in the contribution from 
the nonvalence diagram\cite{BJ}. 
However, the calculation carried out so far was semi-realistic
as the model was 1+1 dimensional and only a point-vertex was considered.

In the present work, we extend our analysis of the fermion-loop to the 
case of 3+1 dimensions. 
In 3+1 dimensions both the covariant and the LF calculations are divergent
and the model without any smeared vertex for the fermion loop is not well 
defined. In the recent literature\cite{Fred,Jaus}, the fermion-loop 
was regulated by smearing the $q\bar{q}$ bound-state vertex in a 
covariant manner. However, the vertex function was not symmetric in the four
momenta of the constituent quarks and could hardly be considered a realistic
approximation of a $q \bar{q}$ bound state. It was regarded 
only as a convenient cutoff prescription which makes the one-loop 
integrals finite\cite{Fred,Jaus}. 
Furthermore, the calculation of the meson decay constant 
reveals that the end-point singularity is not completely cancelled by 
such an asymmetric choice of the vertex function. 
(See the next section for more details.) 
We show in this work that the fermion-loop can also be 
regulated by taking a non-local gauge-boson vertex. 
With this method satisfying the Ward-Takahashi (W-T) identity\cite{WT}, we 
found the complete cancellation of the end-point singularity not only in the
electromagnetic form factor but also in the decay constant.
The non-local gauge-boson vertex remedies also the conceptual 
difficulty associated with the asymmetric way of 
treating $q$ and $\bar{q}$ in the previous 
calculations\cite{Fred,Jaus}. 

Nevertheless, one should distinguish the bound-state from the 
confined-state. In this work, we are treating the mesons as 
bound-states rather than confined-states, because
we do not yet know how to make a covariant regularization for the 
confined-state. Thus, our emphasis here is the inclusion of the 
non-leading order vertex rather than the model-building of a realistic
meson wavefunction. 

We have performed the LF calculation in parallel to the covariant
Feynman calculation. Our light-front results entail the similarity
between the vertex-smearing-technique (either for the bound-state vertex or 
the gauge-boson vertex) and the Pauli-Villars regularization.
For the bosonic loop calculation, the two methods turn out to be identical. 
However, for the fermionic loop calculation, the 
vertex-smearing-technique shares only the same structure of the denominators
with the Pauli-Villars regularization. 
Using the gauge-boson vertex-smearing technique,
we found that the persistent end-point singularity is removed even if the
smeared-vertex is taken to the limit of the point-vertex.

A significant entity of our work is the taxonomy of valence and non-valence
contributions substantiating the finiteness of each contribution when the
gauge-boson vertex is regulated. Our results satisfy current conservation.
However, we note that each contribution individually depends on the 
reference-frame even though the sum is always frame-independent. 
We thus elaborate the frame-dependence of individual contributions.
Also, the zero-mode contribution 
should be distinguished from the instantaneous contribution. For the numerical
estimates of physical observables, we present the electromagnetic form 
factors of the $\pi, K$, and $D$ mesons. 

In the next Section (Section~\ref{sec.2}), we present both the covariant Feynman
calculations and the LF calculations using the LF energy integration
for the electromagnetic form factors of a pseudoscalar meson
with spin-1/2 constituents.
Section~\ref{sec.6} contains the numerical taxonomy
of both the valence and nonvalence diagrams to the 
electromagnetic form factors of $\pi, K$ and $D$-mesons.
The conclusion and discussion follow in Section~\ref{sec.7}.
The general formula including unequal mass cases are summarized in 
Appendix A and the analytic behavior of the valence and nonvalence 
contributions at $Q^2\to0$ limit in various frames is summarized in
Appendix B.
\section{Calculations}
\label{sec.2}

The electromagnetic form factors can be extracted from the matrix
elements of the current $J^\mu$
\begin{equation}\label{eq.01}
\langle p' | J^\mu | p \rangle = i e_m (p^{\prime\mu} + p^\mu) F(q^2),
\end{equation}
where $e_m$ is the charge of the meson and $q^2 = (p'-p)^2$ is the
square of the four momentum transfer. If one uses the plus-component,
$J^+ =J^0 + J^3$, the LF calculation gives two finite
contributions, the {\em LF valence form factor} and the 
{\em LF nonvalence form factor},
that add up to the covariant result, as expected.  The importance of
the nonvalence contribution varies strongly with the momentum
transfer and depends sensitively on the binding energy of the meson.
For small values of $q^2$ and small binding energy, the valence
part is dominant, but elsewhere the nonvalence diagram is essential for
agreement between the LF calculation and the covariant results.

The form factor can also be extracted from the minus-component of the
current, $J^- =J^0 - J^3$. Covariance guarantees that it makes
no difference whether the form factor is determined using the plus or the
minus current matrix element. As LFD is not manifestly covariant, it may
happen that $J^-$ leads to a form factor different from the one determined
using $J^+$.  As we showed in Ref.~\cite{BJ}, the matrix element of $J^-$ 
diverges even in 1+1 dimensional LFD. Unless one regulates $J^-$, the current cannot be 
conserved. To assure current conservation, it was crucial to identify 
the term that causes the divergence.  
We have identified this term exactly and found that it is
an infinite function of the momentum transfer.
If this infinite term is subtracted, the two LF contributions become finite
as it must be in a conserved current.
Moreover, their sum equals again the covariant result as expected.
Still, the regularized LF contributions obtained from $J^-$
are different from the ones
extracted from the plus current. The differences grow with increasing
binding energy.

The covariant fermion triangle-loop (Fig.~\ref{fig0.1}(a)) in 3+1 dimension
is divergent if all the vertices are point-like.
In the recent literature\cite{Fred,Jaus}, the fermion-loop 	
was regulated by smearing the $q\bar{q}$ bound-state vertex 
in a covariant manner. However, the vertex function used in 
Refs.\cite{Fred,Jaus} was not symmetric in the four momenta of the 
constituent quarks as we discussed in the introduction. 
We note that the conceptual difficulty associated with the asymmetry 
could be remedied
if the cutoff prescription is used in the gauge-boson vertex rather than
in the $q\bar{q}$ bound-state vertex. 
In Fig.~\ref{fig0.1}(a), we thus replace the point 
photon-vertex $\gamma^\mu$ by a non-local (or smeared) photon-vertex
$S_\Lambda(k-p) \gamma^\mu S_\Lambda(k-p^\prime)$, 
where $S_\Lambda(p) = \Lambda^2/(p^2 - \Lambda^2 + i\epsilon)$
and $\Lambda$ plays the role of a momentum cut-off. 
Our method is gauge invariant, and satisfies the W-T identity\cite{WT}.
Even though we have computed the covariant 
amplitude (Fig.~\ref{fig0.1}(a))
with unequal constituent masses, for the clarity of presentation
we will focus in this Section on the equal mass case, {\em i.e.} $m_q = 
m_{\bar q} = m$.  The basic formulas for the general case are given in the
Appendix.

The covariant amplitude (Fig.~\ref{fig0.1}(a)) for a 
pseudoscalar meson is given in the equal-mass case by
\begin{equation}\label{eq.02}
\langle p' | J^\mu | p \rangle = 4 N_c\, g^2 \Lambda^4 \int \frac{d^4 k}{(2\pi)^4}
\frac{(m^2-k^2+p\cdot p^\prime) k^\mu + (k^2 - m^2 - k\cdot p^\prime)p^\mu
+(k^2-m^2-k\cdot p) p^{\prime\mu}}{D(k)D(k-p)D(k-p^\prime)D_\Lambda(k-p)
D_\Lambda(k-p^\prime)},
\end{equation}
where $N_c$ is the number of colors and $g$, modulo the obvious charge factor 
$e_m$ is the normalization 
constant fixed by the unity of the form factor at zero momentum transfer
and the denominator factor $D(k)$ from the quark propagator with momentum $k$
is given by $D(k) = k^2 -m^2+i\epsilon$.
While the bound-state vertex is still pointlike, it satisfies 
corresponding relativistic bound-state equation and a binding effect of 
the two-body bound-state is thus included in our analysis.
We note that one can split the denominators in Eq.~(\ref{eq.02}) into
four terms (see below) to show the similarity between the 
methods of photon-vertex-smearing and the Pauli-Villars regularization,
namely the vertex-smearing-technique shares the same structure of the 
denominators with the Pauli-Villars regularization.
To investigate the issue of the end-point singularity\cite{BJ}, 
we present the $J^-$ calculation in the following. 

First, the valence contribution shown in Fig.~1(b) is obtained
in the range $0 < k^+ < p^+$. By picking up the pole of the 
spectator quark, {\it i.e.} 
$k^- = \frac{m^2 + {\vec k_\perp}^2 -i\epsilon}{k^+}$, we obtain
\begin{eqnarray}\label{eq.jminus}
\langle p' | J^- | p \rangle &=& 
\frac{-4\pi i N_c\, g^2 \Lambda^4}{(\Lambda^2 - m^2)^2}
\int\frac{dk^+ d^2{\vec k_\perp}}{(2\pi)^4} 
\frac{\{-k^+ p^- {p^\prime}^- 
+{\vec k_\perp}\cdot{\vec {p^\prime}_\perp} p^- 
+{\vec k_\perp}\cdot{\vec p_\perp} {p^\prime}^- 
- {\vec p_\perp}\cdot{\vec {p^\prime}_\perp} k^-\}}
{k^+ (k^+ - p^+)(k^+ - {p^\prime}^+)} \nonumber\\
&&\times\left[\frac{1}{E(p,\Lambda)E(p^\prime,\Lambda)}
-\frac{1}{E(p,\Lambda)E(p^\prime,m)}
- \frac{1}{E(p,m)E(p^\prime,\Lambda)} 
+\frac{1}{E(p,m)E(p^\prime,m)} \right],\nonumber\\
\label{eq2.3}
\end{eqnarray} 
where the energy denominator $E(p,\Lambda)$ is defined as
\begin{equation}\label{eq.val}
E(p,\Lambda) = \frac{m^2 + {\vec k_\perp}^2}{k^+} - p^- - \frac{\Lambda^2 
+ ({\vec k_\perp} - {\vec p_\perp})^2}{k^+ - p^+}.
\end{equation}
As we see from Eq.~(\ref{eq2.3}), the result depends on the reference frame.
However, as we will see later  the sum of the 
valence and nonvalence contributions will of course be identical in
any frame. In the present Section, we choose the frame where 
$q^+=\alpha p^+$ and ${\vec p_\perp}= 0$ and the momentum fraction $x$ is 
defined by $k^+ =(1-x)p^+$. 
Then the LF valence form factor (see Eq.~(\ref{eq.01})) is given by
\begin{eqnarray}\label{eq.Fval}
F^-_{\rm val}(q^2) &=& 
\frac{2 N_c\, g^2 \Lambda^4}{(2+\alpha)+{\vec q_\perp}^2 / M^2} 
\int_0^1 dx \int \frac{d^2{\vec k_\perp}}{(2\pi)^3} 
\frac{(1-x)(M^2+{\vec q_\perp}^{\,2}) 
- (1+\alpha){\vec k_\perp}\cdot{\vec q_\perp}}
{(1-x)x^2(x+\alpha)^2} \nonumber \\
&\times&\frac{1}{
\left(\frac{m^2+{\vec k_\perp}^2}{1-x} +
\frac{m^2+{\vec k_\perp}^2}{x}-M^2\right)
\left(\frac{m^2+{\vec k_\perp}^2}{1-x} + 
\frac{\Lambda^2+{\vec k_\perp}^2}{x}-M^2\right)} \nonumber \\
&\times&\frac{1}{
\left(\frac{m^2+{\vec k_\perp}^2}{1-x}+\frac{m^2+({\vec k_\perp}-{\vec 
q_\perp})^2}{x+\alpha}-\frac{M^2+{\vec q_\perp}^{\, 2}}{1+\alpha}\right)
\left(\frac{m^2+{\vec k_\perp}^2}{1-x}+\frac{\Lambda^2+({\vec k_\perp}-{\vec 
q_\perp})^2}{x+\alpha}
-\frac{M^2+{\vec q_\perp}^{\, 2}}{1+\alpha}\right)},
\end{eqnarray}
where $M$ is the meson mass.
The limit to the point vertex can be taken by letting $\Lambda\to\infty$
and we find that the result is finite in this frame.

Next, the nonvalence contribution shown in Fig.~1(c) is obtained
in the range $p^+ < k^+ < {p^\prime}^+$. Here, the pole is not taken
as the spectator but as $k^- = {p^\prime}^- + \frac{m^2 + ({\vec 
k_\perp} -{\vec {p^\prime}_\perp})^2 - i\epsilon}{k^+ - {p^\prime}^+}$
(for the term corresponding to the Pauli-Villars particle, $k^- = 
{p^\prime}^- + \frac{\Lambda^2 + ({\vec 
k_\perp} -{\vec {p^\prime}_\perp})^2 - i\epsilon}{k^+ - {p^\prime}^+}$).
Following a procedure similar to the one described in the valence case,
we find the LF nonvalence form factor to be given by
\begin{eqnarray}\label{eq.NV}
F^-_{\rm nv}(q^2) &=& - \frac{2 N_c\, g^2 \Lambda^4 (1+\alpha)}{(\Lambda^2 - m^2)
[M^2 (2+\alpha) + {\vec q_\perp}^{\, 2}] } \nonumber \\
&&\times \int_0^\alpha dx \int \frac{d^2{\vec k_\perp}}{(2\pi)^3}
\frac{1}{x^2 (1+x)(\alpha-x)} 
\left[ \frac{N(m)}{D0 D1 D2}
- \frac{N(\Lambda)}{ D5 D3 D4}\right],
\end{eqnarray}
where
\begin{eqnarray}\label{eq.001}
 N(m) & = &\left[ m^2+{\vec k_\perp}^2 - (1+x)\left(\frac{M^2 + {\vec 
q_\perp}^2}{1+\alpha}-\frac{m^2+({\vec k_\perp}-{\vec 
q_\perp})^2}{\alpha-x}\right)\right] \left[\frac{m^2 + ({\vec k_\perp}-{\vec 
q_\perp})^2}{\alpha-x}+M^2 \right] \nonumber \\
 & & +\frac{(1+x)M^2(M^2+{\vec 
q_\perp}^2)}{1+\alpha}-{\vec k_\perp}\cdot{\vec q_\perp} M^2,
 \nonumber \\
 N(\Lambda) & = &\left[ m^2+{\vec k_\perp}^2 - (1+x)\left(\frac{M^2 + {\vec 
q_\perp}^2}{1+\alpha}-\frac{\Lambda^2+({\vec k_\perp}-{\vec q_\perp})^{\, 2}} 
{\alpha-x}\right)\right] \left[\frac{\Lambda^2 + ({\vec k_\perp}-{\vec 
q_\perp})^2}{\alpha-x}+M^2 \right] \nonumber \\
 & & +\frac{(1+x)M^2(M^2+{\vec q_\perp}^{\, 2})}{1+\alpha}
 -{\vec k_\perp}\cdot{\vec q_\perp} M^2,
 \nonumber \\
 D0 & = & \frac{M^2 + {\vec q_\perp}^2}{1+\alpha}
-\frac{m^2+({\vec k_\perp}-{\vec q_\perp})^{\, 2}}{\alpha-x}
-\frac{m^2 + {\vec k_\perp}^2}{1+x},
\nonumber \\
 D1 & = & M^2-\frac{M^2 + {\vec q_\perp}^2}{1+\alpha}
+\frac{m^2+({\vec k_\perp}-{\vec q_\perp})^{\, 2}}{\alpha-x}
+\frac{\Lambda^2 + {\vec k_\perp}^2}{x},
 \nonumber \\
 D2 & = & M^2-\frac{M^2 + {\vec 
q_\perp}^2}{1+\alpha}+\frac{m^2+({\vec k_\perp}-{\vec q_\perp})^2}{\alpha-x}
+\frac{m^2 + {\vec k_\perp}^2}{x},
 \nonumber \\
 D5 & = & \frac{M^2 + {\vec q_\perp}^{\, 2}}{1+\alpha}-
 \frac{\Lambda^2+({\vec k_\perp}-{\vec q_\perp})^2}{\alpha-x}
 -\frac{m^2 + {\vec k_\perp}^2}{1+x},
 \nonumber \\
 D3  & = & M^2-\frac{M^2 + {\vec q_\perp}^2}{1+\alpha}
+\frac{\Lambda^2+({\vec k_\perp}-{\vec q_\perp})^2}{\alpha-x}
+\frac{\Lambda^2 + {\vec k_\perp}^2}{x},
 \nonumber \\
 D4 & = & M^2-\frac{M^2 + {\vec q_\perp}^2}{1+\alpha}
+\frac{\Lambda^2+({\vec k_\perp}-{\vec q_\perp})^2}{\alpha-x} 
+\frac{m^2 + {\vec k_\perp}^2}{x}.
\end{eqnarray}
We find from Eq.~(\ref{eq.NV}) that the two terms in the integrand
are individually finite as long as the parameter $\Lambda$ is finite.
There is neither an ultra-violet divergence nor an end-point singularity.
However, in the limit to the point vertex ({\it i.e.} $\Lambda\to\infty$),
we find that each term has not only a linear divergence in the ${\vec 
k_\perp}$ integration, but also an end-point singularity in the $x$ 
integration, which cancel each other exactly. Thus, in the point vertex limit,
we find that the end-point singularity is completely removed even though
the result is logarithmically divergent as it must be in the 3+1 dimensional
fermion-loop with point vertices. 
This shows a striking difference from the calculation without relying on the
vertex regularization from the beginning\cite{BJ}.
The critical reason for this is that the end-point singularity for 
the fermion-loop
is a consequence of the bottomless nature of the Dirac sea and
the vertex-smearing effectively provides the weighting in the Dirac sea
de-emphasizing the lower part. The previously
identified end-point singularity\cite{BJ} is exactly cancelled by the 
identical end-point singularity from the term generated by the 
vertex-smearing corresponding to the Pauli-Villars particle. 
Therefore, in the regularized case, 
all the physical degrees of freedom are taken into account.

In addition, the sum of the valence and nonvalence contributions 
$F(q^2) = F_{\rm val}(q^2) + F_{\rm nv}(q^2)$ is of course identical to 
the results obtained by other components of the current, {\it i.e.}
either $J^+$ or $J^\perp$\cite{Fred,Jaus}.\footnote{In the electromagnetic
form factor, there is no zero-mode contribution in $J_{\perp}$ current.}
Also, the net result $F(q^2)$ is independent of the choice of 
reference frame. 

A similar calculation can be made for the pseudoscalar-meson decay constant 
$f$. Taking a non-local gauge-boson vertex, we verified again the exact 
cancellation of linear and logarithmic divergences in the (two-point) 
fermion loop. The result for the equal mass case such as the pion is 
given by 
\begin{equation}
f = \frac{N_c\, g\, m \Lambda^4}{4\sqrt{2}\pi^2(\Lambda^2-m^2)^2}\int_0^1 dx 
 \log\left[\frac{C(m,m)C(\Lambda,\Lambda)}{C(m,\Lambda)C(\Lambda,m)}\right],
\label{eq.decay}
\end{equation}
where $C(m_1,m_2)$ is given by
\begin{equation}
C(m_1,m_2)=x(1-x)M^2-(1-x)m_1^2-xm_2^2.
\label{eq.C}
\end{equation}
If the asymmetric $q\bar{q}$ bound-state vertex-smearing is used
as in Refs.\cite{Fred,Jaus}, the above result, Eq.(\ref{eq.decay}),
is replaced by
\begin{equation}
f = \frac{N_c\, g\, m \Lambda^2}{4\sqrt{2}\pi^2(\Lambda^2-m^2)}\int_0^1 dx 
 \log\left[ \frac{C(m,m)}{C(m,\Lambda)} \right].
\label{eq.decay-asym}
\end{equation}
We note that the logarithmic divergence is not completely 
cancelled in the unequal mass case if the minus component of the weak 
current is used with the asymmetric $q\bar{q}$ bound-state vertex-smearing.
The non-local gauge-boson vertex used in this work does not suffer from 
such an incomplete cancellation of divergences no matter which
component of the current is used.

In the next section, we present the numerical
taxonomy of the $\pi$, $K$, and $D$ meson 
electromagnetic LF form factors choosing various reference-frames,
as well as the values of the decay constants $f_\pi, f_K$ and $f_D$. 

\section{Numerical Results}
\label{sec.6}
\begin{table}
\caption{Our model parameters of constituent quark masses $m$, and 
cutoff values $\Lambda$ (in units of GeV) used in this work. 
The decay constants $f^{\rm sym}$ [$f^{\rm asym}$] obtained 
by Eq.~(\ref{eq.decay}) [Eq.~(\ref{eq.decay-asym})] are also compared
with the experimental values $f^{\rm exp}$~\protect\cite{PDG}.\label{tab1}}
\begin{tabular}{cccccccc}
Meson & $m_q$&  $\Lambda_q$ & $m_{\bar{q}}$ & $\Lambda_{\bar{q}}$
& $f^{\rm sym}[f^{\rm asym}]$ & $f^{\exp}$[MeV] \\
\hline
$\pi$ & 0.25 &0.90  & 0.25 & 0.90 & 92.5  [122.2]  
& 92.4 $\pm$0.25 \\
$K$   & 0.25 &0.90  & 0.48 & 0.91 & 112.5 [139.4]  
& 113.4$\pm$1.1 \\
$D$   & 1.78 &1.79  & 0.25 & 0.90 & 108.6 [191.5]  
& $\leq 154.9$
\end{tabular}
\end{table}

In Table~\ref{tab1}, we present our model parameters such as
the constituent quark masses ($m$), and the cutoff values ($\Lambda$), 
as well as the decay constants that we calculated here and compared to the 
experimental data~\cite{PDG}. 
The meson masses ($M$) are taken as the experimental
values~\cite{PDG}. Our model parameters have been chosen to fit both the 
charge radii and the decay constants to the experimental data well, 
although the available data for the decay constant of the $D$-meson is 
only an upper limit. We also compare in Table~\ref{tab1} the decay 
constants from our symmetric non-local gauge-boson vertex ($f^{\rm sym}$)
with those from the asymmetric $q\bar{q}$ bound-state vertex 
($f^{\rm asym}$) smearing case.
However, we note that our calculation here is limited in value because
the zero-range approximation is used for the bound-state vertices.
\begin{table}
\caption{The kinematics in the reference frames used in this work, where
$\kappa=Q^2/2M$ and $\hat{n}=(\cos\phi,\sin\phi)$.\label{tab2}}
\begin{tabular}{cccc}
Kinematics & Target rest frame& Breit frame& DYW frame\\
\hline
 $q^{+}$ & $\kappa+Q\sqrt{1+\kappa/2M}\cos\theta$ & $+Q\cos\theta$ & 0 \\
 $q^{-}$ & $\kappa-Q\sqrt{1+\kappa/2M}\cos\theta$ & $-Q\cos\theta$ &
 $Q^2/p^+$ \\
 $\vec{q}_{\perp}$ & $Q\sqrt{1+\kappa/2M}\sin\theta\hat{n}$
 & $Q\sin\theta\hat{n}$ & $Q\hat{n}$ \\
\hline
 $p^+$ & $M$ & $\sqrt{M^2 + Q^2/4}-q^+/2$ & $p^+$ \\
 $p^-$ & $M$ & $\sqrt{M^2 + Q^2/4}-q^-/2$ & $M^2/p^+$ \\
 $\vec{p}_{\perp}$ & 0 & $-\vec{q}_{\perp}/2$ & 0
\end{tabular}
\end{table}

Three different reference frames were considered: the 
Drell-Yan-West (DYW) frame, the target-rest frame (TRF)
and the Breit frame. The corresponding kinematics is
given in Table~\ref{tab2}.
The DYW frame has gained some popularity in deep-inelastic
scattering calculations because in that frame $q^+ = 0$ identically.
This frame can be obtained by taking the $\alpha\to0$ limit in the frame
presented in section II. In the limit $\alpha\to1$, the frame presented
in section II coincides with the target-rest frame with $\theta=0$.
 
Now we comment on our results shown in the figures below. (In all these
figures, we use thick solid line for the covariant form factor, thick
dashed line for $F^{+}_{\rm val}$, thick dot-dashed line for 
$F^{+}_{\rm nv}$, thin dashed and dot-dashed lines for the corresponding
minus LF form factors.) We show in Fig.~\ref{fig0.2} the results of our 
numerical calculations using the DYW frame and compare with the experimental 
data~\cite{Brown,Brauel,Ackerman,Amen3,Volmer,Amen2}
Our total results, represented by the solid line in each figure,
are also in very good agreement with the experimental data of the pion 
and kaon form factors, respectively. 

In the DYW frame $F^+_{\rm nv}$ vanishes
identically. Remarkably, we find that the nonvalence part of the minus
current, which in this reference frame coincides with the zero-mode
contribution, makes a very important contribution to the total form factor 
and may even dominate over the valence part in the whole $Q^2$-range
considered. There are quantitative differences between the results obtained
for the different mesons, which are to large extent due to the difference in
binding: the tighter the binding, the more important $F^-_{\rm nv}$ becomes.
We checked the binding effect in the case of the pion. By varying the
quark mass alone from the realistic value of $0.25$ GeV to $0.07007$ GeV, 
thus lowering the binding energy to 0.1\% of the pion mass, we found that 
the value of $F^-_{\rm val}$ at $Q^2 = 0$ was increased from 6\% to 69\%.
Still, for $Q^2 > 0.3$ GeV$^2$ $F^-_{\rm val}$ is lower than $F^-_{\rm nv}$.
This indicates that for the larger values of $Q^2$, the relativistic effects 
can still be large.

The frame dependence of the different components of the current can be
studied by comparing the results of calculations in different frames and at
different values of the polar angle $\theta$. It is worth mentioning that
the results must be independent of the azimuthal angle $\phi$, because
rotations about the $z$-axis are kinematical transformations.
(We used this fact as a check on the correctness of our codes.)

In Fig.~\ref{fig0.3} we show the results of our numerical
calculations in the Breit frame at $\theta = 0$, which differs from the
target-rest frame at $\theta = 0$ by a boost in the $z$-direction only,
so the results are identical in these two frames. The first thing we notice
is the great difference with the DYW frame. Now there is a 
sizeable contribution from $F^+_{\rm nv}$, which dominates at higher values 
of the momentum transfer: at $Q^2=0$, $F^+_{\rm val}$ coincides with the 
covariant form factor. 
It crosses $F^+_{\rm nv}$ at some value of $Q^2$, the crossover
point being smaller for larger binding energy. 
It is of special interest to separate the instantaneous part, i.e.
the contribution to $F^+$ from diagrams with one internal
instantaneous propagator. They are given in the figures by crosses
($\times$).  It turns out that they give a very large contributions to the
pion form factor but become negligible for the heavy $D$-meson case.

Turning to the minus current we see that in this reference frame the relative
importance of $F^-_{\rm nv}$ becomes more prominent than in the DYW
frame for the more tightly bound mesons $\pi$ and $K$. In the case of the pion 
the dominance of $F^-_{\rm nv}$ is so strong, that it can hardly be 
distinguished from the covariant form factor.
The figures (in Fig.~\ref{fig0.3}) suggest that at $Q^2 = 0$ the values 
obtained are frame independent. This is indeed the case, as we found in our 
calculations and can be understood as the equality of the form factors in 
the long-wave-length limit. 

A first glance at the angle dependence is given in Fig.~\ref{fig0.4}.
In both the target-rest frame and the Breit frame for
$\theta = \pi/2$ the transferred momentum is purely transverse. As the
pure Lorentz boosts in the transverse direction, $K_1$ and $K_2$,
are dynamical in LFD, we must expect a strong frame
dependence and this is indeed what we find. As $q^+ = 0$ in the Breit frame
for $\theta = \pi/2$, the part $F^+_{\rm nv}$ vanishes identically.
Still, the DYW frame results differ for the minus current, which is 
clear from a comparison with Fig.~\ref{fig0.2}. The reason is that
they are not connected by a kinematical transformation. For the same
reason the zero-mode contributions differ in the two frames. In the
DYW frame it is very close to the total form factor, while in the
Breit frame it even overshoots the covariant form factor at large values of
$Q^2$ by a factor of almost 2.
In this connection we want to mention the work of Frederico
et al.~\cite{Fred}, who performed a calculation similar in spirit to ours, 
but claimed that in the Breit frame at $\theta=\pi/2$ $F^-_{\rm val}$ is 
always very close to the covariant form factor. 
This discrepancy was due to the difference of their $F^-_{\rm val}$ 
definition in the Breit frame at $\theta = \pi/2$, where they removed
the term that is odd under the transformation $p^- - k^- \rightarrow
-(p^- - k^-)$ in their definition\cite{Fredtrick}.
However, our definition of $F^-_{val}$ is general for any angle $\theta$
and our results on $\theta$-dependence are 
smooth as shown in Fig.~\ref{fig0.13}.

We show the systematics of the angle dependence for the case of the pion only.
Our results are depicted in Figs.~\ref{fig0.5}-\ref{fig0.8}. ( The thick solid
line for $\theta=0$, thin dotted line for $\theta=\pi/4$, the thick dashed
line for $\theta=\pi/4$, the thin dot-dashed for $\theta=3\pi/4$, and the
the thick long-dashed line for $\theta=\pi$.) One sees
immediately that the angle dependence is smooth but can be very strong, both
for the valence and the nonvalence parts, calculated from either the plus
or the minus current.

One might try to exploit the angle dependence to optimize the calculation of
the form factor in a noncovariant framework. However, as the figures above
clearly show, there is no value for the angle $\theta$ where both
$F^+_{\rm nv}$ and $F^-_{\rm nv}$ are negligible, or even suppressed,
compared to the valence parts, for all values of $Q^2$.  On the contrary, as 
the values of the form factor components at $Q^2 = 0$ are frame independent, 
we can be sure that $F^-_{\rm nv}$ must be very important for an important 
part of the $Q^2$ range.

As a summary of what we found concerning the angle dependence we show in
Fig.~\ref{fig0.13} the complete angle dependence for two values of the
momentum transfer, $Q^2=0$ and 0.1 GeV$^2$. All curves in this figure are
clearly smooth and demonstrate the fact that there is no preferred value for
the polar angle $\theta$.

\section{Conclusion and Discussion}
\label{sec.7}

In this paper, we have analyzed all the components of
the current quantized on the light-front to compute the electromagnetic
form factors of pseudoscalar mesons with spin-1/2
consituents. 
Since our aim in this work was to analyze the taxonomy of the triangle 
diagram, we did not choose any particular electromagnetic gauge but just 
presented the equivalence of the physical form factor $F(q^2)$ in any choice
of the current component.
The divergence appearing in the 3+1 dimensional 
fermion-loop 
calculations was regulated by the covariant vertex-smearing-technique.
Performing the light-front calculation, we verified that the 
vertex-smearing-technique is similar to the Pauli-Villars regularization.
In the $J^-$ computation, we find that
the critical smearing effect persists even in the limit to the point vertex
because the end-point singularity existing otherwise is completely removed
once the limit is taken at the end of the calculation. If the limit is
taken at the beginning of the calculation, however, we have alreay shown
that the end-point singularity in the nonvalence contribution leads to an 
infinitely different result from that obtained by the covariant Feynman 
calculation~\cite{BJ}. Our taxonomical analysis demonstrated that
each individual contribution, whether valence or nonvalence, is finite regardless
of which component of the current is considered. However, we stress that each contribution
depends on the reference-frame even though the sum does not.
Of course, the invariance of the sum ensures the current 
conservation. Also, the zero-mode contribution should be distinguished
from the instantaneous contribution as we have numerically estimated
the differences.

>From our numerical calculations we can conclude that to obtain agreement
with the covariant form factor one needs both the valence and the nonvalence 
parts. For tightly bound states the nonvalence parts dominate in an 
important part of the range of $Q^2$-values that we studied. 
It is natural that this result
runs counter to nonrelativistic intuition, which says that the valence
parts should dominate, because the tightly bound states are not expected 
to be non-relativistic.
In Ref.~\cite{BJ} it was demonstrated in the
1+1-dimensional case, that by weakening the binding $F^+_{\rm val}$ and
$F^-_{\rm val}$ approximate the covariant form factor more and more  closely.
Here we found that indeed for less tightly bound state the valence parts
come closer to the covariant result, but even at $Q^2 =0$ $F^-_{\rm val}$
gives only 69\% of the covariant form factor.

If two reference frames can be connected by a kinematical Lorentz 
transformation the LF form factors calculated in these two frames must 
be the same.
Otherwise they must differ. We found that the angle dependence within a
chosen reference frame is always smooth, although it may be very strong.
For some values of the polar angle $\theta$ two reference frames may be
connected by a kinematical Lorentz transformation, e.g., the target-rest frame
and the Breit frame for $\theta = 0$. On the other hand, 
the DYW frame and the  Breit frame for $\theta =0$
are not connected by a kinematical transformation and the
form factor components consequently do not coincide.
More details of the frame-dependence can be found in Ref.~\cite{ChJ}.

While the calculations carried out in this work is more realistic 
than the case of 1+1 dim with the point vertex, they are still
semi-realistic as the model uses a bound-state rather than a confined-state.
Moreover, we used for the vertices those obtained from a Bethe-Salpeter
equation using a contact interaction\cite{SM} which are of zero range
and may emphasize the importance of the nonvalence contributions rather
differently from a more realistic model.
Thus, there is still much room for extending our model towards a more 
realistic model. However, the essential conclusions
about the similarity to the Pauli-Villars regularization and the 
cancellations of both ultra-violet divergence and the end-point singularity 
remain intact. Nevertheless, the numerical results may differ
from a more realistic model calculation. 
This point is presently under investigation.

\acknowledgements
\noindent
This work was supported in part by a grant from the US Department of
Energy and the National Science Foundation. 
This work was started when HMC and CRJ visited the Vrije Universiteit
and they want to thank the staff of the department of physics at 
VU for their kind hospitality.
BLGB wants to thank the staff of the department of physics 
at NCSU for their warm hospitality during a stay when this work was 
completed. The North Carolina Supercomputing
Center and the National Energy Research Scientific Computer Center are also
acknowledged for the grant of Cray time.
\newpage
\appendix

\section{Unequal-mass case}

If we number the momenta of the internal lines of the fermion triangle
as $k_1$ (spectator), $k_2=p + k_1$, and $k_3 = p' + k_1$ (struck quark) 
respectively, and the
corresponding masses as $m_1$, $m_2$, and $m_3$ then we find for the
trace appearing in the numerator of the covariant integral, the expression
\begin{equation}
 T^\mu
 = 4  \left[ -(m_2 m_3 - \vec{k}_{2\, \perp} \cdot \vec{k}_{3\, \perp}) k^\mu_1
 + (m_1 m_3 - \vec{k}_{2\, \perp} \cdot \vec{k}_{3\, \perp}) k^\mu_2
 + (m_1 m_2 - \vec{k}_{2\, \perp} \cdot \vec{k}_{3\, \perp}) k^\mu_3 \right].
\label{eq0.01}
\end{equation}
The valence diagram is obtained if one calculates the integral over
$k^-_1$ by closing the contour around the pole corresponding to putting the
spectator $k_1$ on shell. This corresponds to the following values for 
the minus-components of the momenta to be used in the expression for $T^\mu$
\begin{equation}
 k^-_1 = \frac{\vec{k}^2_{1\,\perp} + m^2_1}{k^+_1}, \quad
 k^-_2 = p^- + k^-_1, \quad k^-_3 = p^{\prime\, -} + k^-_1.
\label{eq0.02}
\end{equation}
To obtain the nonvalence diagram one closes the contour around the
pole corresponding to $k_3$. Then one gets for the part of the
propagator with the mass $m_3$
\begin{equation}
  k^-_3 =  \frac{\vec{k}^2_{3\,\perp} + m^2_3}{k^+_3}, \quad
  k^-_1 = -p^{\prime\, -} + k^-_3, \quad k^-_2 = p^- - p^{\prime\, -} + k^-_3.
\label{eq0.03}
\end{equation}
For the part of the propagator with the cutoff $\Lambda'$ the same
formula can be used, but with $m_3$ replaced by $\Lambda'$. 

As the smearing we use affects only the denominator parts of the
propagator of the struck quark, the replacement of $m_3$ by $\Lambda'$
in $T^\mu$ occurs only in the minus components. 

The energy denominators are easy to find. One obtains for the valence parts
\begin{eqnarray}
 D(k-p) & = & p^- + \frac{\vec{k}^2_{1\,\perp} + m^2_1}{k^+_1}
 - \frac{\vec{k}^2_{2\,\perp} + m^2_2}{k^+_2},
 \nonumber \\
 D(k-p') & = & p^{\prime\, -} + \frac{\vec{k}^2_{1\,\perp} + m^2_1}{k^+_1}
			  - \frac{\vec{k}^2_{3\,\perp} + m^2_3}{k^+_3}.
\label{eq0.04}
\end{eqnarray}
The nonvalence part has the same denominator $D(k-p')$, but $D(k-p)$ is 
changed to
\begin{equation}
 D'(k-p) = p^- - p^{\prime\, -} + \frac{\vec{k}^2_{3\,\perp} + m^2_3}{k^+_3}
 - \frac{\vec{k}^2_{2\,\perp} + m^2_2}{k^+_2}.
\label{eq0.04a}
\end{equation}
The energy denominators $D_\Lambda$ and $D'_{\Lambda'}$ are obtained by
the substitutions $m_2 \to \Lambda$ and $m_3 \to \Lambda'$, respectively.

In the case where $m_1 = m_2 = m_3$ and $\Lambda = \Lambda'$ the 
final formulas are much simplified. 
The explicit formulas in the main text are valid for the equal-mass case.
The general case can be easily
constructed from the expressions (\ref{eq0.01} - \ref{eq0.04a}).

In the explicit formulas we use the notation
\begin{equation}
 k^{\mu} = - k^{\mu}_1, \quad x =  \frac{k^+}{p^+} 
\; ({\rm valence}), \; x = \frac{k^+ - p^+}{p^+} 
 \; ({\rm nonvalence}), \quad \alpha = \frac{p^{\prime\, +}- p^+}{p^+}.
\label{eq0.05}
\end{equation}

\section{Analyticity}

>From the covariant expression for the amplitude one can prove that the
form factors are analytic functions of $Q^2$. This proof is not valid
for the LF-time-ordered amplitudes. One can, however, expand the
expressions that one obtains for the different parts of the form factor
in terms of $Q$ and determine the $Q$-dependence at small values of
$Q^2$. It turns out that the results depend on the kinematics: it
matters in which reference frame one does the calculations. 

We use the fromulas for the momenta given in Table~\ref{tab2}
and expand the trace and the energy denominators in powers of
$Q$. From this expansion the analyticity properties of the
amplitudes follow. 
We have numerically verified that a blow-up of 
Figs.~\ref{fig0.5}-\ref{fig0.8} for small
$Q^2$ illustrates the analyticity properties of the amplitudes
discussed in this Appendix.
The three different reference frames are discussed
consecutively.

\subsection{Target-rest frame}
We discuss the valence part of the plus-current in detail;
the other LF form factors can be treated in a similar way.
First we expand the momenta. 
\begin{eqnarray}
 q^\pm & = & q \cos \theta + {\cal O} (Q^2), \quad
 \vec{q}_\perp = Q \sin \theta \, \hat{n} + {\cal O} (Q^3), \nonumber \\
 p^{\prime\, \pm} & = & M + Q \cos \theta + {\cal O} (Q^2), \quad
 \vec{p}^{\,\prime}_\perp = Q \sin \theta \, \hat{n} + {\cal O} (Q^3).
\label{eq0.07}
\end{eqnarray}

The trace consists of a piece that is independent of $Q$ and a piece
that consists of two parts, one that is proportional to 
$Q \vec{k}_{1\, \perp} \cdot \hat{n} \sin \theta$
and a part proportional to $Q \cos \theta$.
The denominator $D(k-p')$ has a similar behaviour, but
$D(k-p)$ is independent of $Q$. 

Upon integration over $\vec{k}_{1\, \perp}$ the terms proportional to 
$\vec{k}_{1\, \perp} \cdot \hat{n} \sin \theta$ vanish. Consequently,
the valence part $F^+_{\rm val}$ has the small-$Q$ behaviour
\begin{equation}
 F^+_{\rm val} \sim F^{+\, 0}_{\rm val} + F^{+\, 1}_{\rm val} Q \cos \theta .
\label{eq0.08}
\end{equation}

The other cases, nonvalence plus-current, valence and nonvalence minus-current,
show the same pattern. There are pieces independent of $Q$, parts with the
$Q$-dependence $Q  \vec{k}_{1\, \perp} \cdot \hat{n} \sin \theta$ 
and ones proportional to $Q \cos \theta$. As the pieces proportional 
to $\vec{k}_{1\, \perp} \cdot \hat{n}$ vanish upon integration, all the
components of the form factor show a behaviour similar to 
Eq.~(\ref{eq0.08}). So, only for $\theta = \pi/2$ do we find that the LF 
time-ordered amplitudes calculated in the target-rest frame are analytic 
in $Q^2$.

\subsection{Breit frame}

In the case of the Breit frame we can follow the same line as in the case of the
target-rest frame. The only difference is that now no terms of the form 
$Q \vec{k}_{1\, \perp} \cdot \hat{n} \sin \theta$ appear. As those terms give
no contribution to first order in $Q$ anyway, this does not alter the result:
also in the Breit frame the components of the form factor have the
same small-$Q$ behaviour as in Eq.~(\ref{eq0.08}).

\subsection{Drell-Yan-West frame}

In the Drell-Yan-West frame there is no angle dependence. However, there is
a term linear in $Q$. It is proportional to 
$Q \hat{n} \cdot \vec{k}_{1\, \perp}$. Of course, this term
also vanishes upon integration over $\vec{k}_{1\, \perp}$. 
Therefore, in the Drell-Yan-West frame we find no term linearly dependent 
on $Q$, so the amplitudes are analytic in $Q^2$.

\subsection{Zero mode}
The zero mode is defined as the contribution to the nonvalence amplitude 
that survives the limit $q^+ \to 0$. It is easy to see that only the 
minus-current can have a zero-mode part, because the integral defining 
$F^+_{\rm nv}$ has an integrand that remains finite when $q^+$ goes to 0. 
However, the integrand defining $F^-_{\rm nv}$ diverges when the limit
$q^+ \to 0$ is taken. In order to determine the limit, one may expand the 
integrand in powers of $q^+$. As $q^+$ is taken to be zero in the 
Drell-Yan-West frame, the analysis cannot be done in that frame, but it can 
most easily be carried out in the Breit frame.

The algebra being straightforward but tedious, we shall not give the details. 
Rather we quote the final result. If we consider the smeared case, the most 
divergent part of the integrand for $F^-_{\rm nv}$ has the behaviour 
$1/q^{+\, 2}$. As the integration over $k^+_1$ ranges from $-p^{\prime\, +}$ 
to $-p^+$, we can scale the integration variable as in Eq.~(\ref{eq0.05}) 
and obtain an integral over $x$ from 0 to 1. The Jacobian being $q^+$, it 
cancels one factor $q^+$ in the denominator, so we only need to show
that the leading term vanishes to prove that the zero-mode contribution
is finite. 

Upon carrying out the algebra we find the following result. 
The leading part is proportional to
\begin{equation}
 x^2 (\vec{k}^2_{3\,\perp} + m^2_3) (\vec{k}^2_{3\,\perp} 
+ \Lambda^{\prime\, 2})
 - (1-x)^2 (\vec{k}^2_{2\,\perp} + m^2_2) (\vec{k}^2_{2\,\perp} + \Lambda^2).
\label{eq0.09}
\end{equation}
This part of the integrand does not vanish, because 
$\vec{k}_{2\, \perp} = \vec{p}_\perp - \vec{k}_{\perp}
\neq \vec{k}_{3\, \perp} = \vec{p}^{\,\prime}_\perp - \vec{k}_{\perp}$.
However, if we take $\Lambda' = \Lambda$ and $m_2 = m_3$, being the mass of 
the struck quark, 
and as $\vec{k}_{2\, \perp}$, $\vec{k}_{3\, \perp}$, and
$\vec{k}_{\perp}$ differ only by a constant vector, this function 
vanishes after integration over $x$ from 0 to 1 
and $\vec{k}_{\perp}$ over the whole of ${\sf I\! R}_2$. 
So we see that the coefficient of the contribution proportional to $1/q^+$ 
vanishes. The remaining part, the piece that survives the limit
$q^+ \to 0$, is the zero mode contribution. In the Breit frame as well as 
the target-rest frame it is well-defined and finite.

\newpage
\begin{figure}
\begin{center}
 \epsfig{file=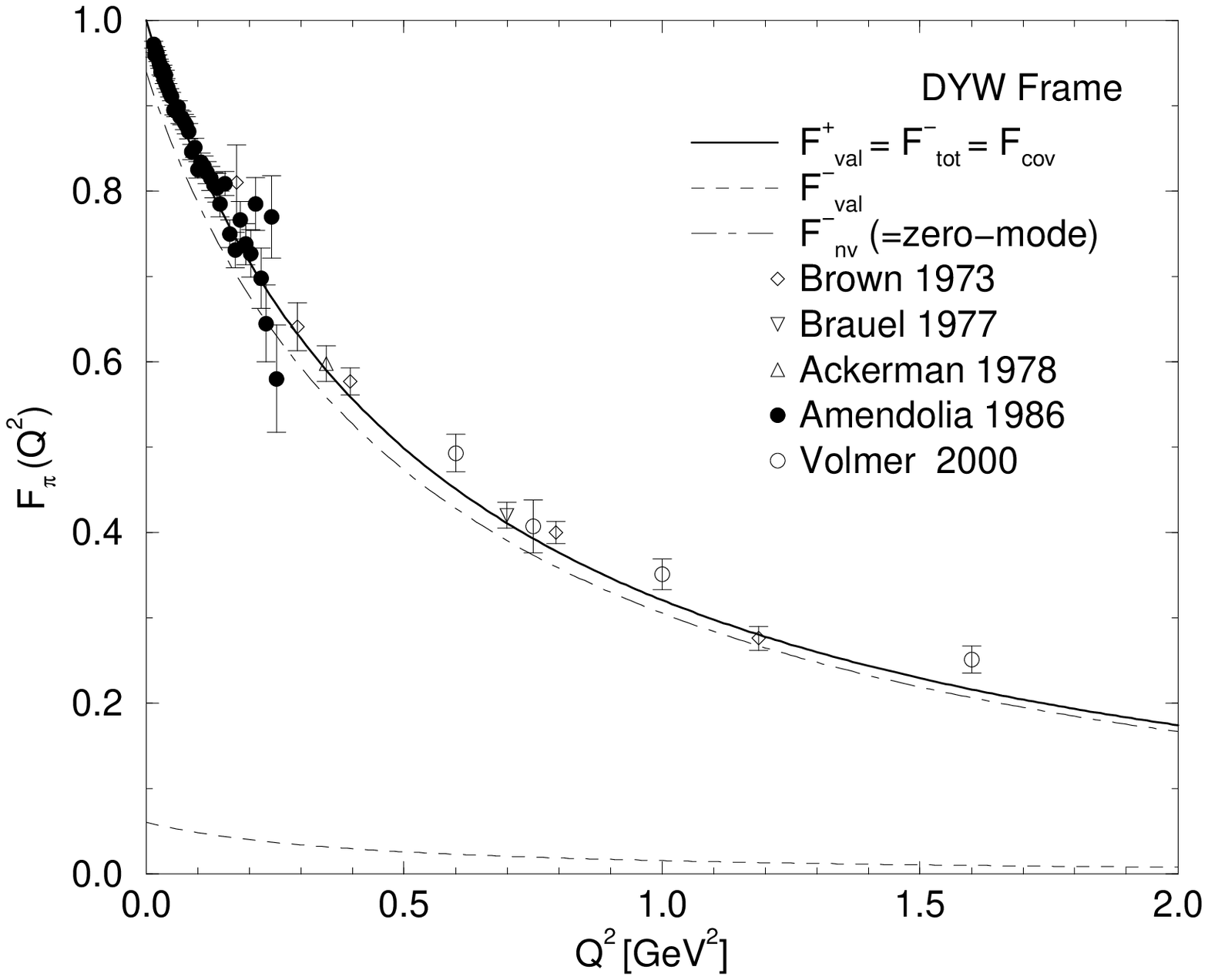,height=69mm,width=69mm} \\
 \epsfig{file=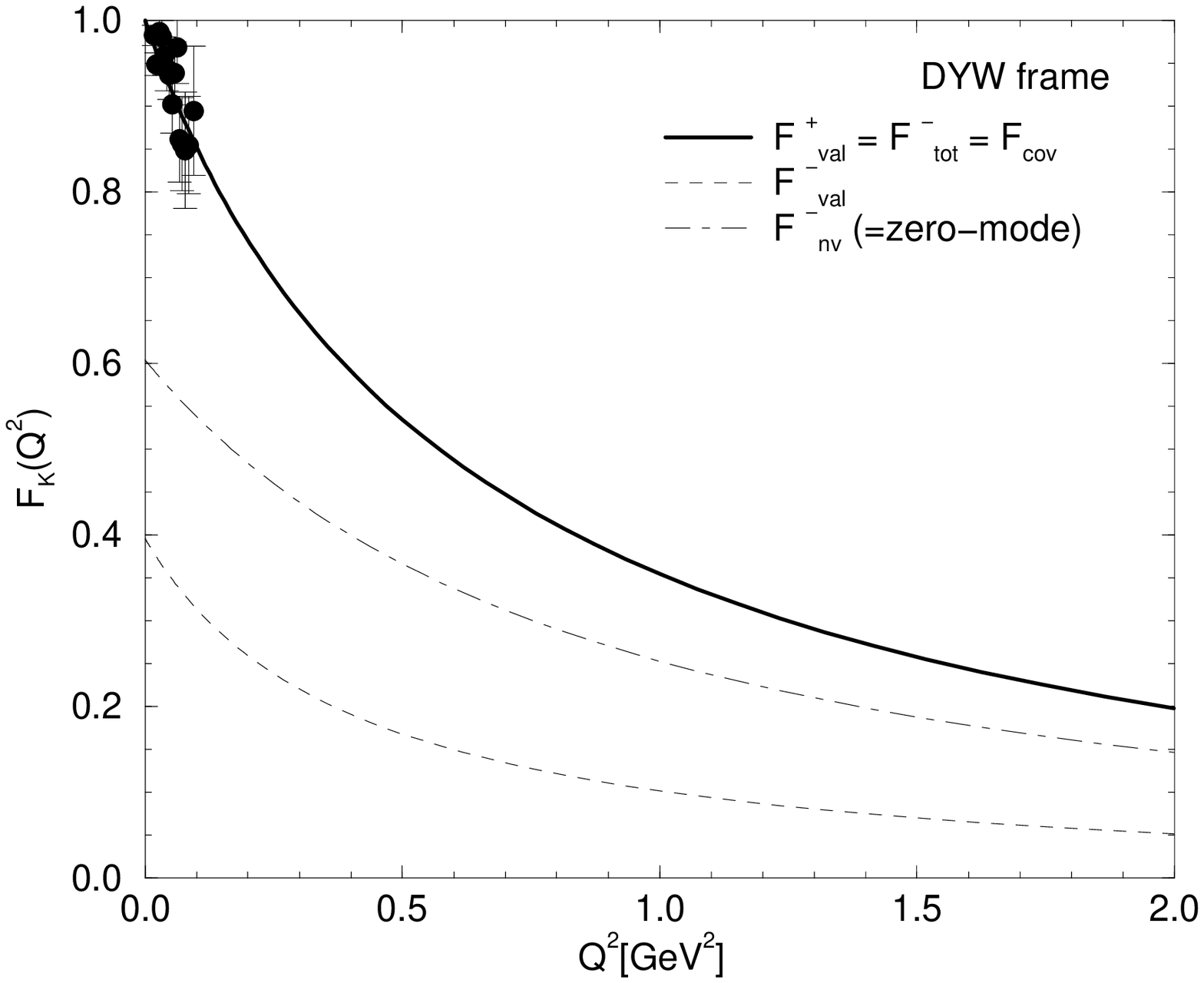,height=80mm,width=80mm} 
 \epsfig{file=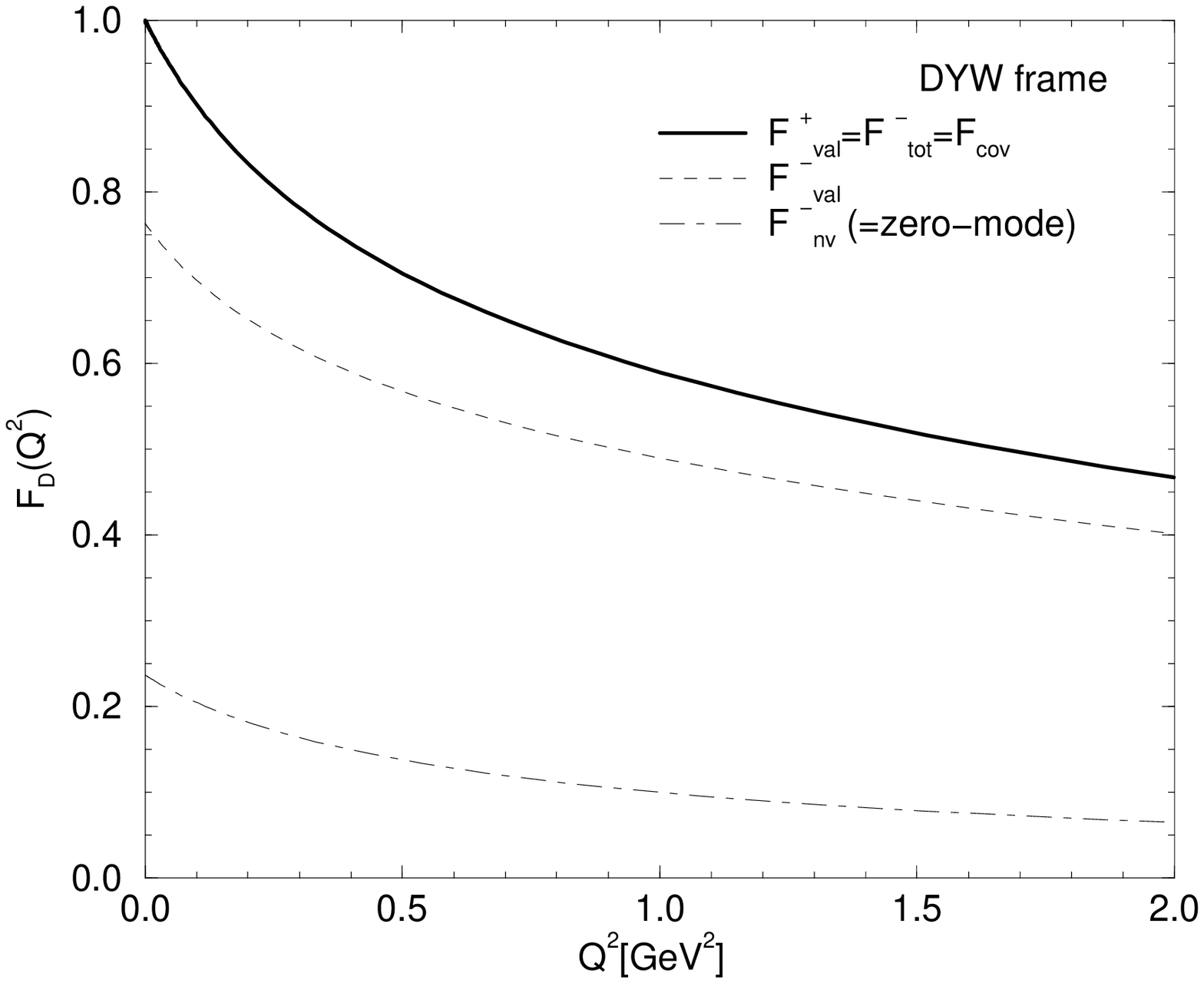,height=80mm,width=80mm}
\caption{Pion, $K$, and $D$ meson form factors in the Drell-Yan-West (DYW)
frame compared with experimental data for 
pion~\protect\cite{Brown,Brauel,Ackerman,Amen3,Volmer} and kaon~\protect\cite{Amen2}. The part $F^+_{\rm nv}$
 vanishes identically in this reference frame. \label{fig0.2}}
\end{center}
\end{figure}
\begin{figure}
\begin{center}
 \epsfig{file=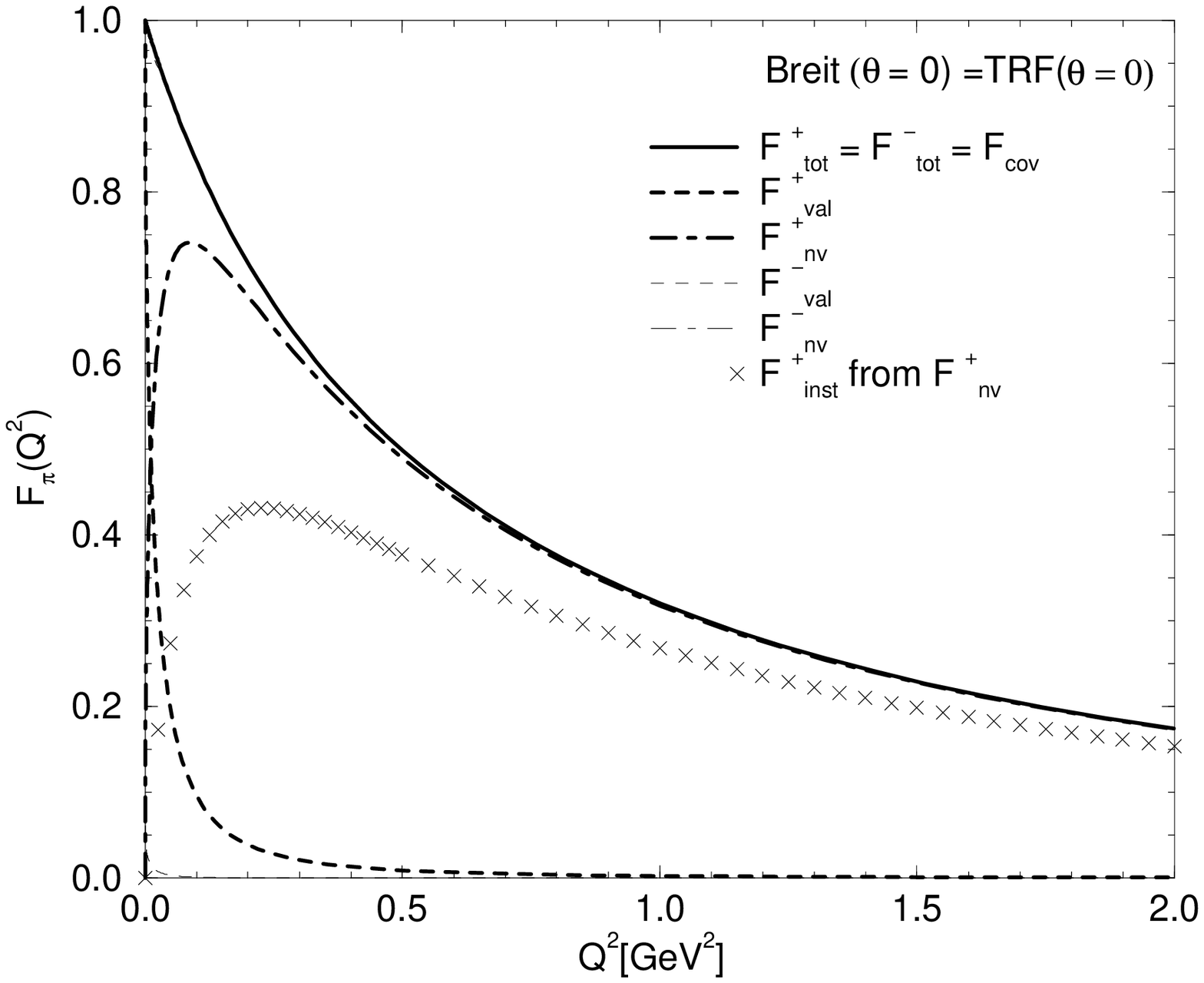,height=80mm,width=80mm}
 \epsfig{file=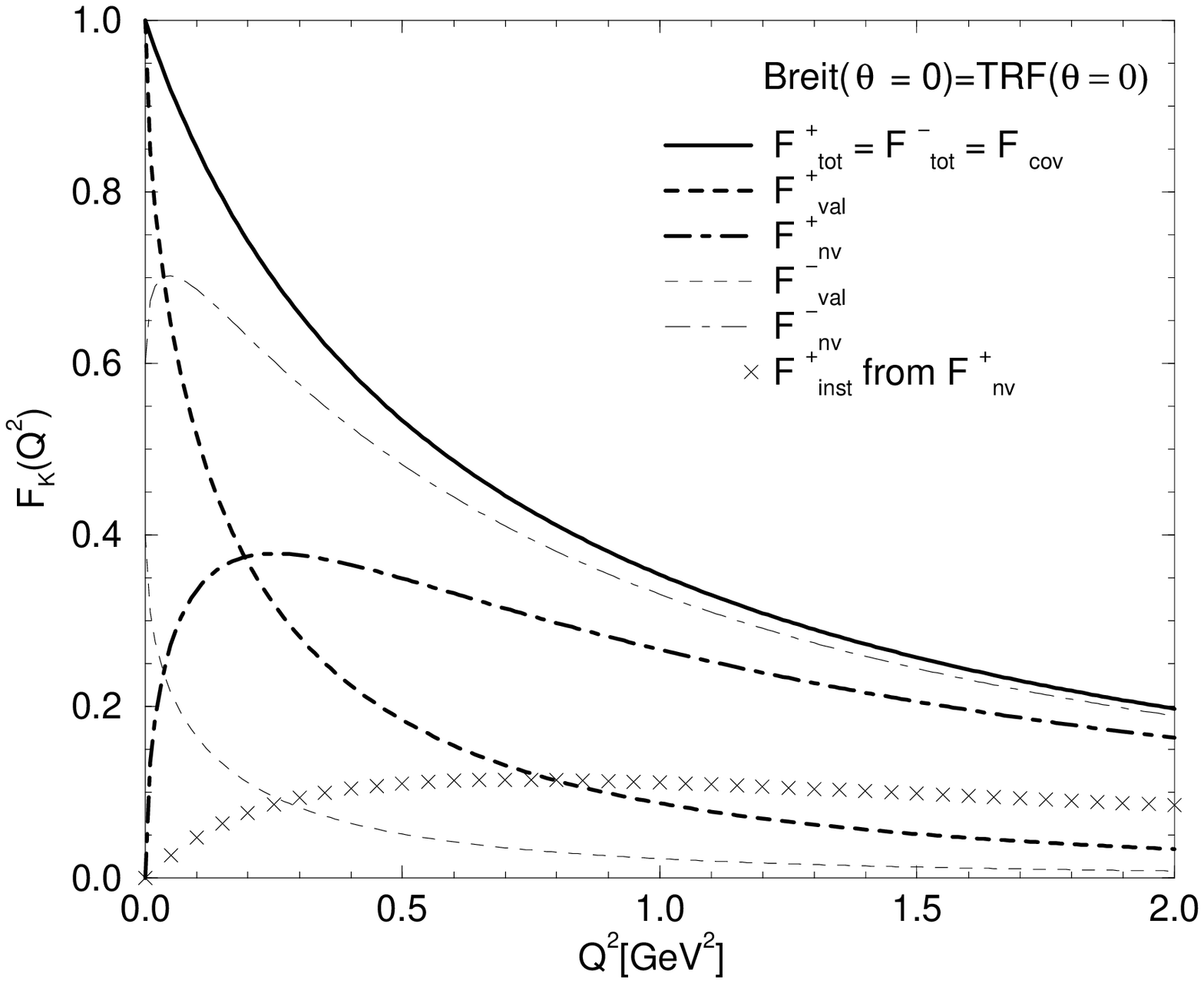,height=80mm,width=80mm}
 \epsfig{file=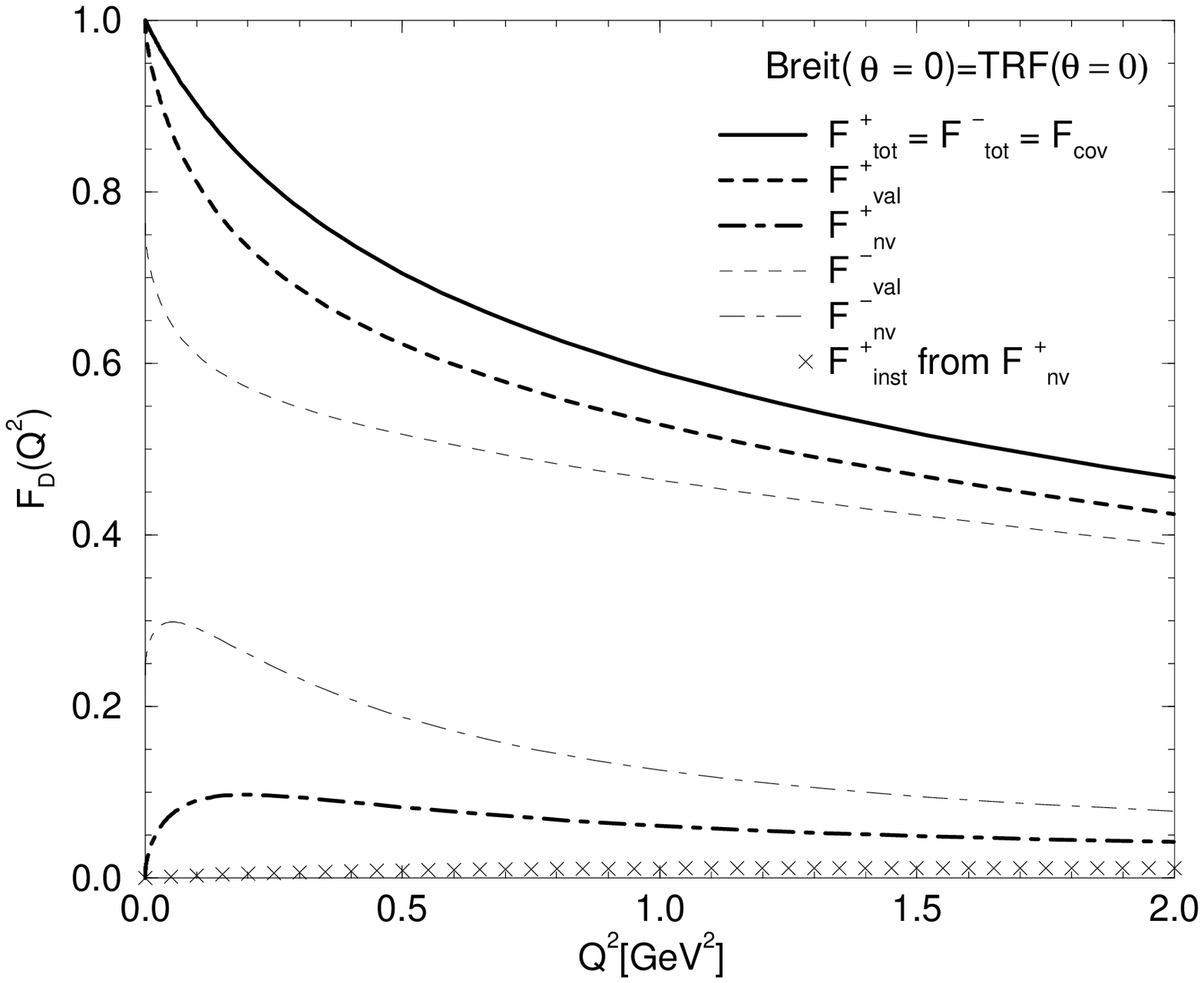,height=80mm,width=80mm}
\caption{Pion, $K$, and $D$ meson form factors in the target-rest 
frame (TRF) or the Breit frame, both at $\theta = 0$. \label{fig0.3}}
\end{center}
\end{figure}
\begin{figure}
\begin{center}
 \epsfig{file=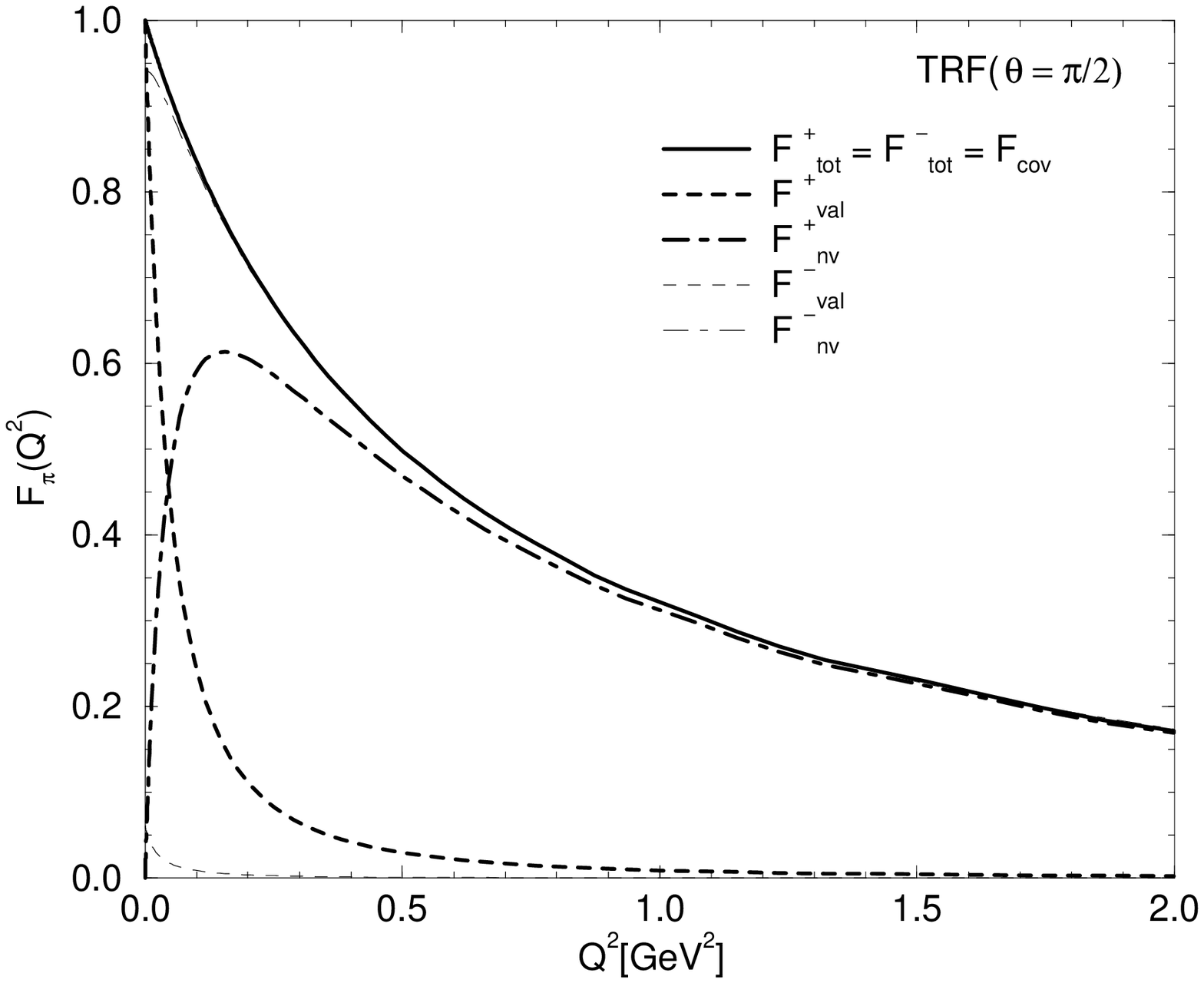,height=80mm,width=80mm}
 \epsfig{file=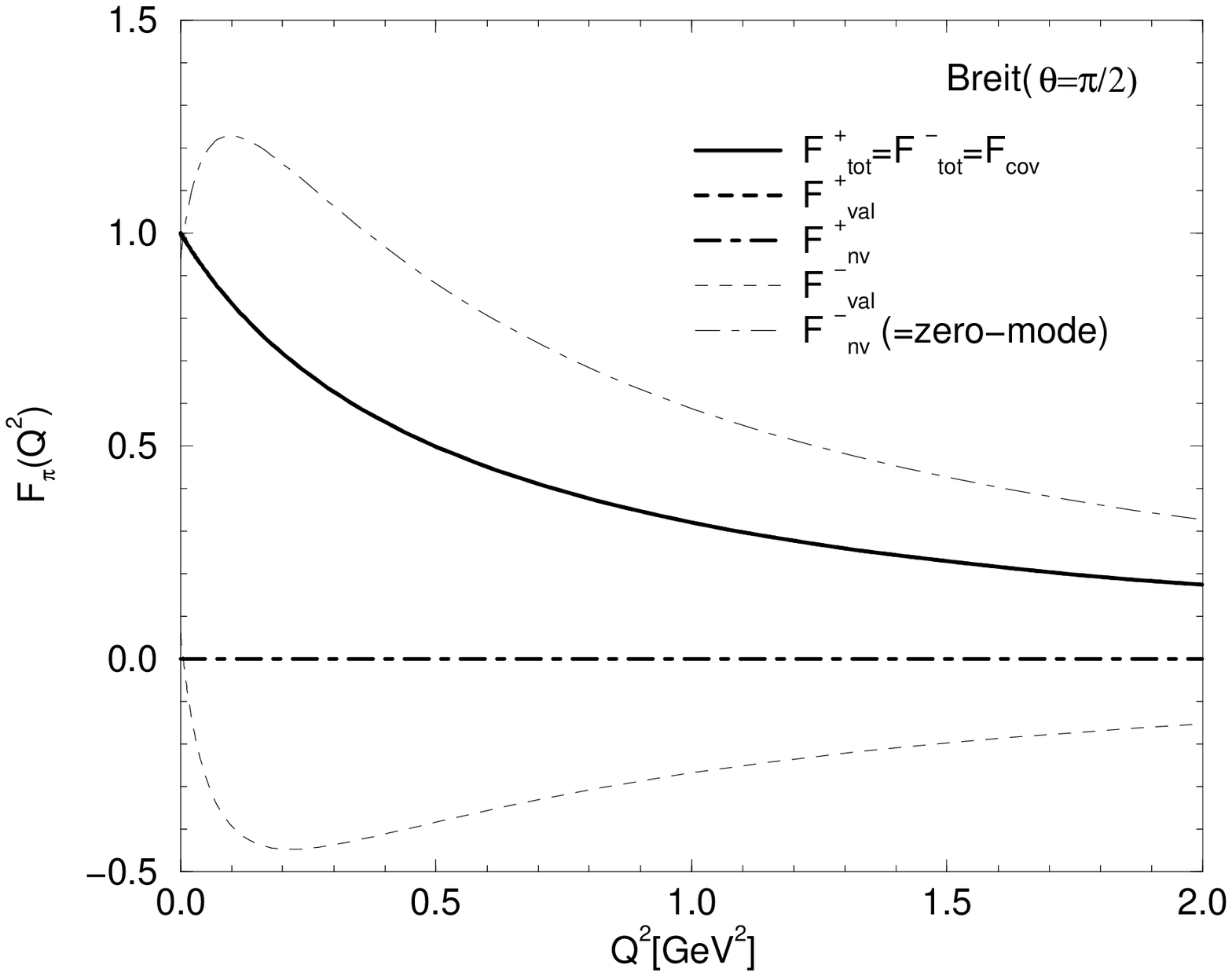,height=80mm,width=80mm}
\caption{Pion form factor in the target-rest frame and the Breit frame
 at $\theta=\pi/2$. \label{fig0.4}}
\end{center}
\end{figure}
\begin{figure}
\begin{center}
 \epsfig{file=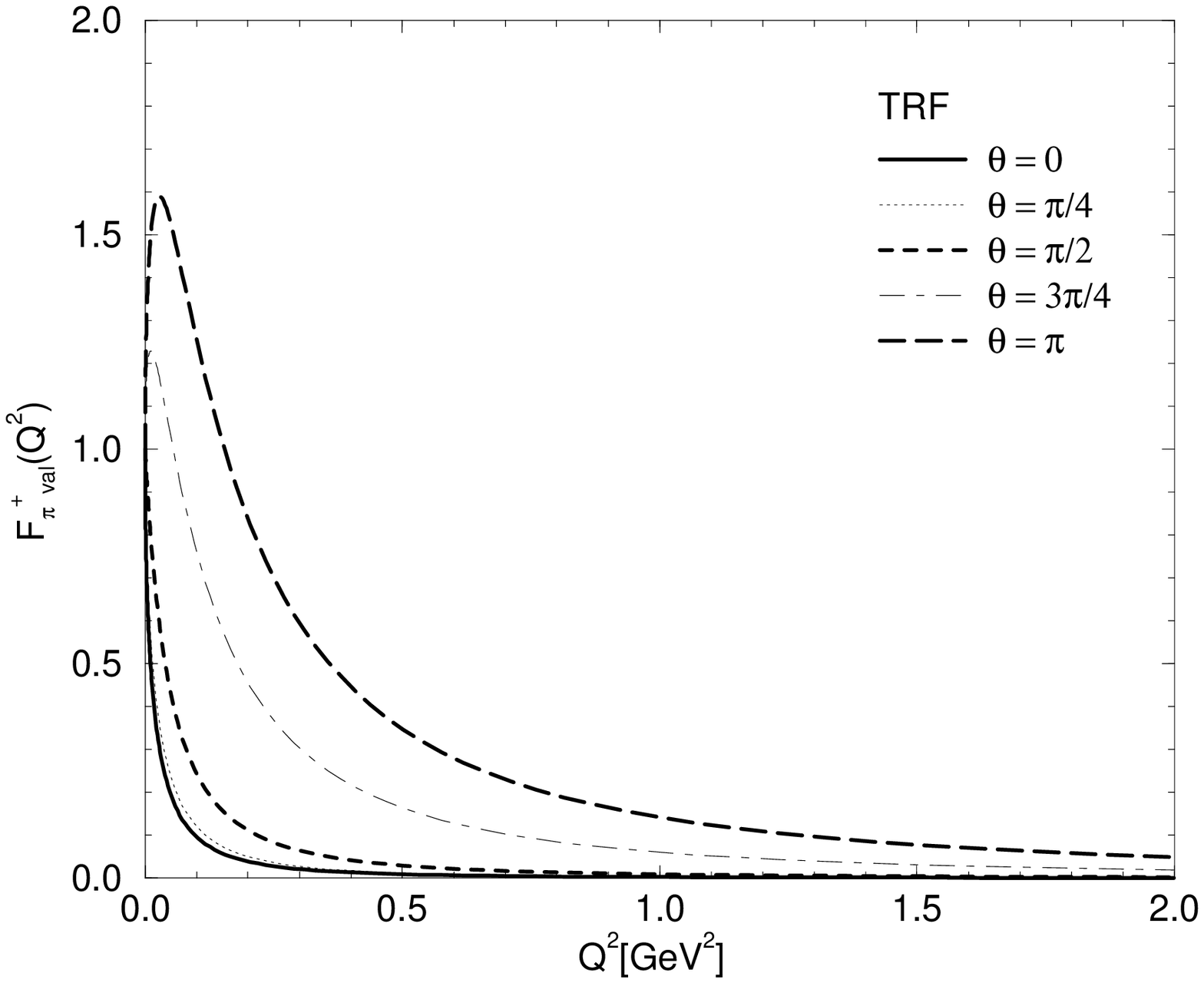,height=80mm,width=80mm}
 \epsfig{file=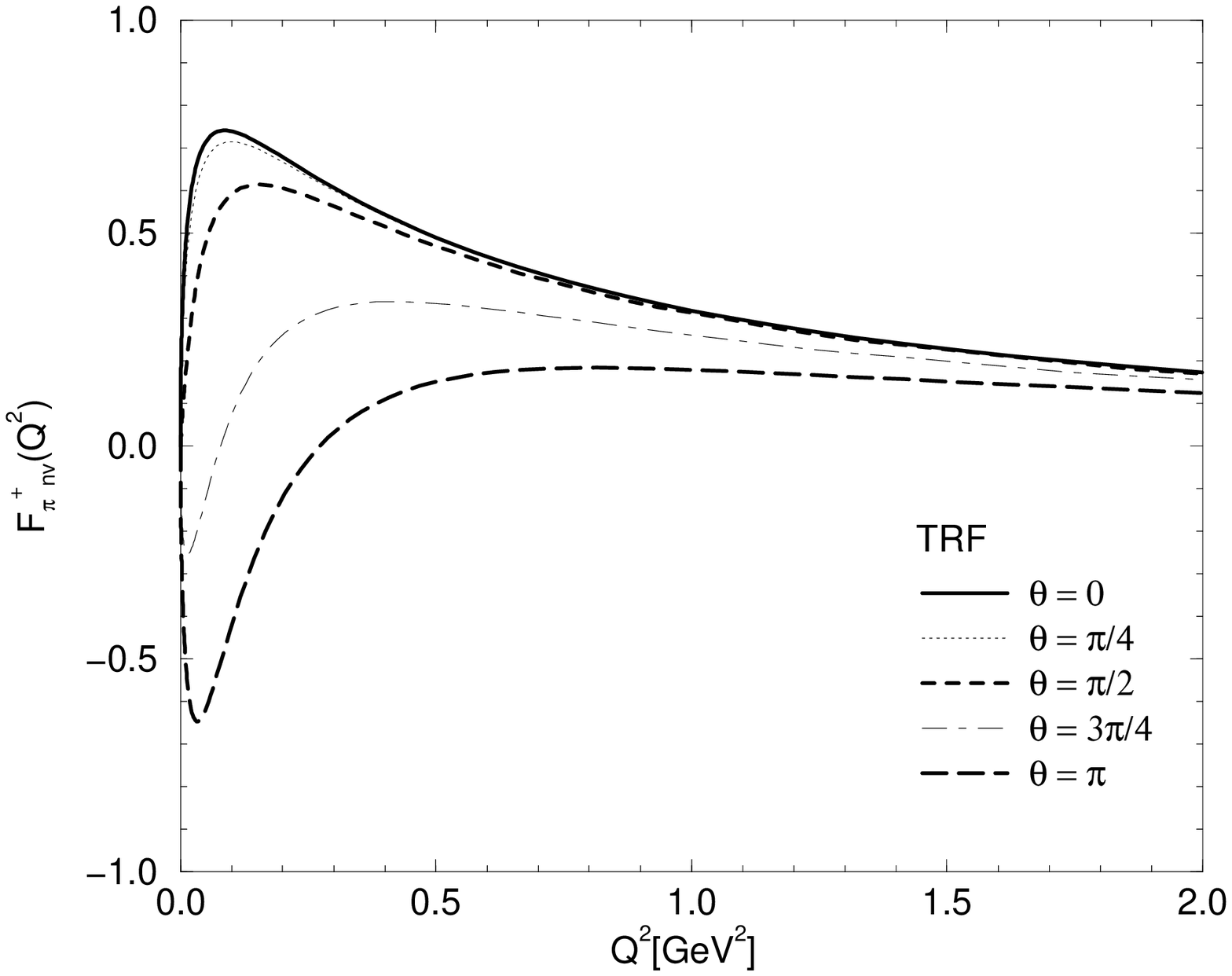,height=80mm,width=80mm}
\caption{Pion LF form factors $F^+_{\rm val}$ and $F^+_{\rm nv}$
 in the target-rest frame for five different
 values of the polar angle $\theta$. \label{fig0.5}}
\end{center}
\end{figure}
\begin{figure}
\begin{center}
 \epsfig{file=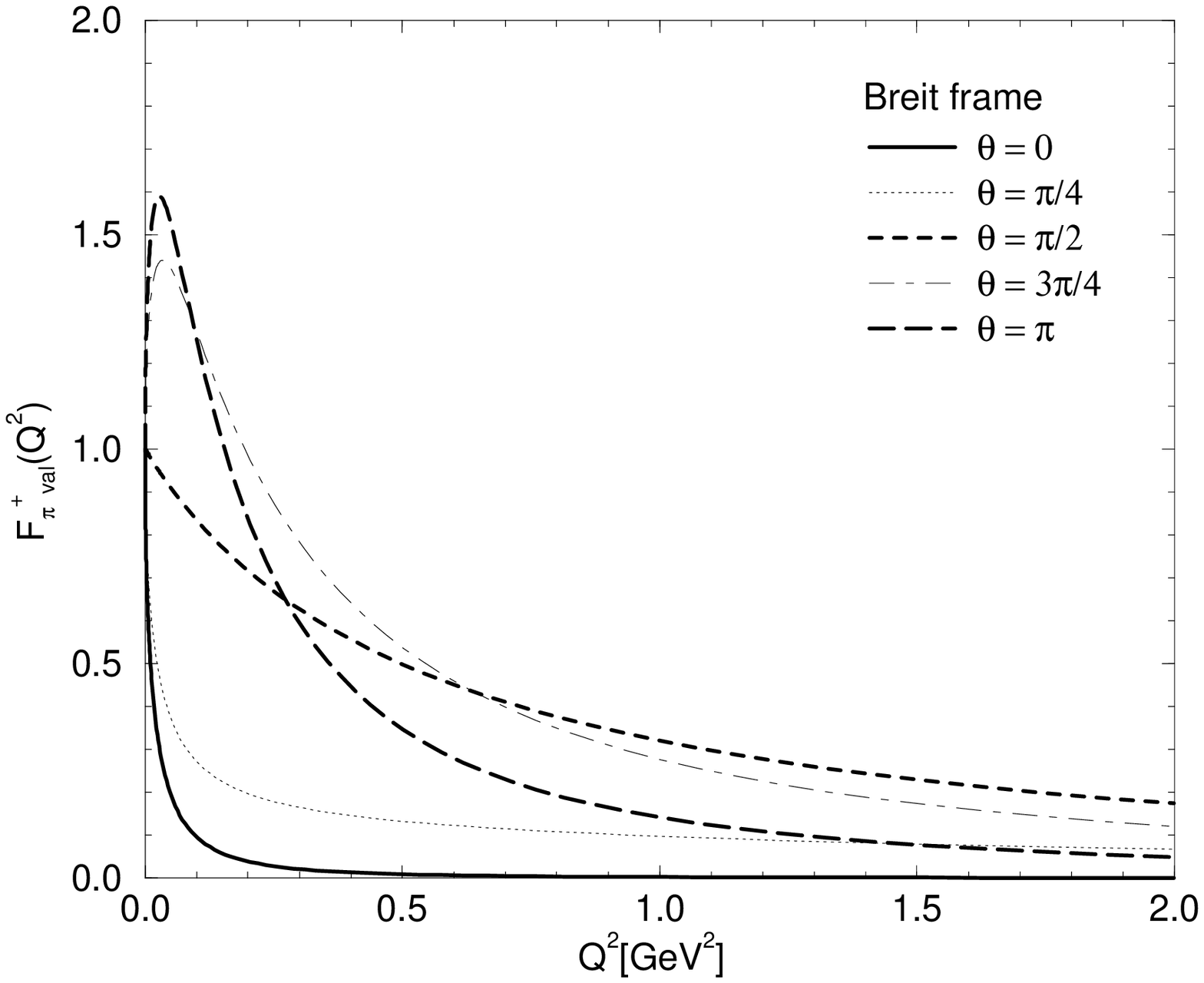,height=80mm,width=80mm}
 \epsfig{file=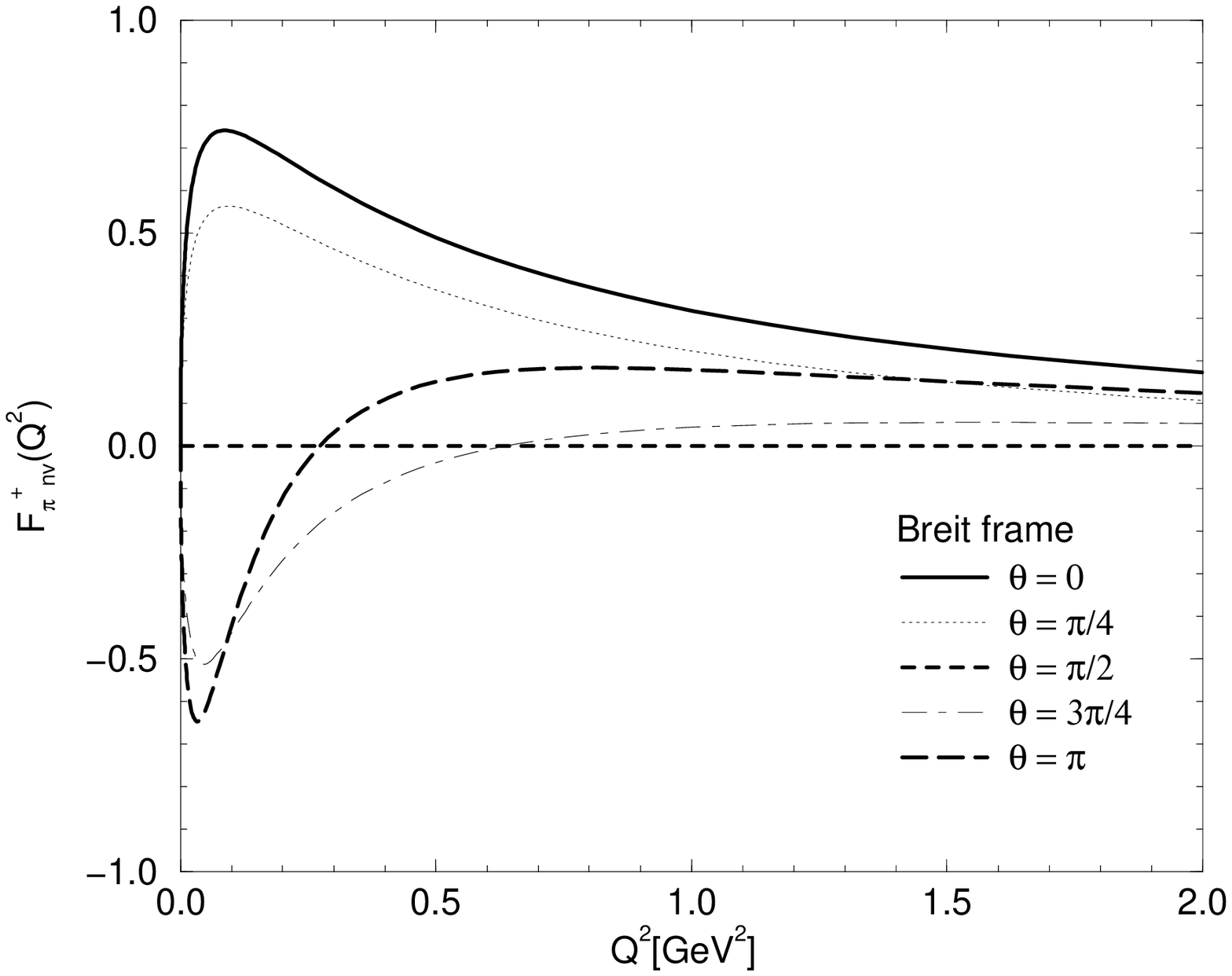,height=80mm,width=80mm}
\caption{Pion LF form factors $F^+_{\rm val}$ and $F^+_{\rm nv}$
 in the Breit frame for five different
 values of the polar angle $\theta$. \label{fig0.6}}
\end{center}
\end{figure}
\begin{figure}
\begin{center}
 \epsfig{file=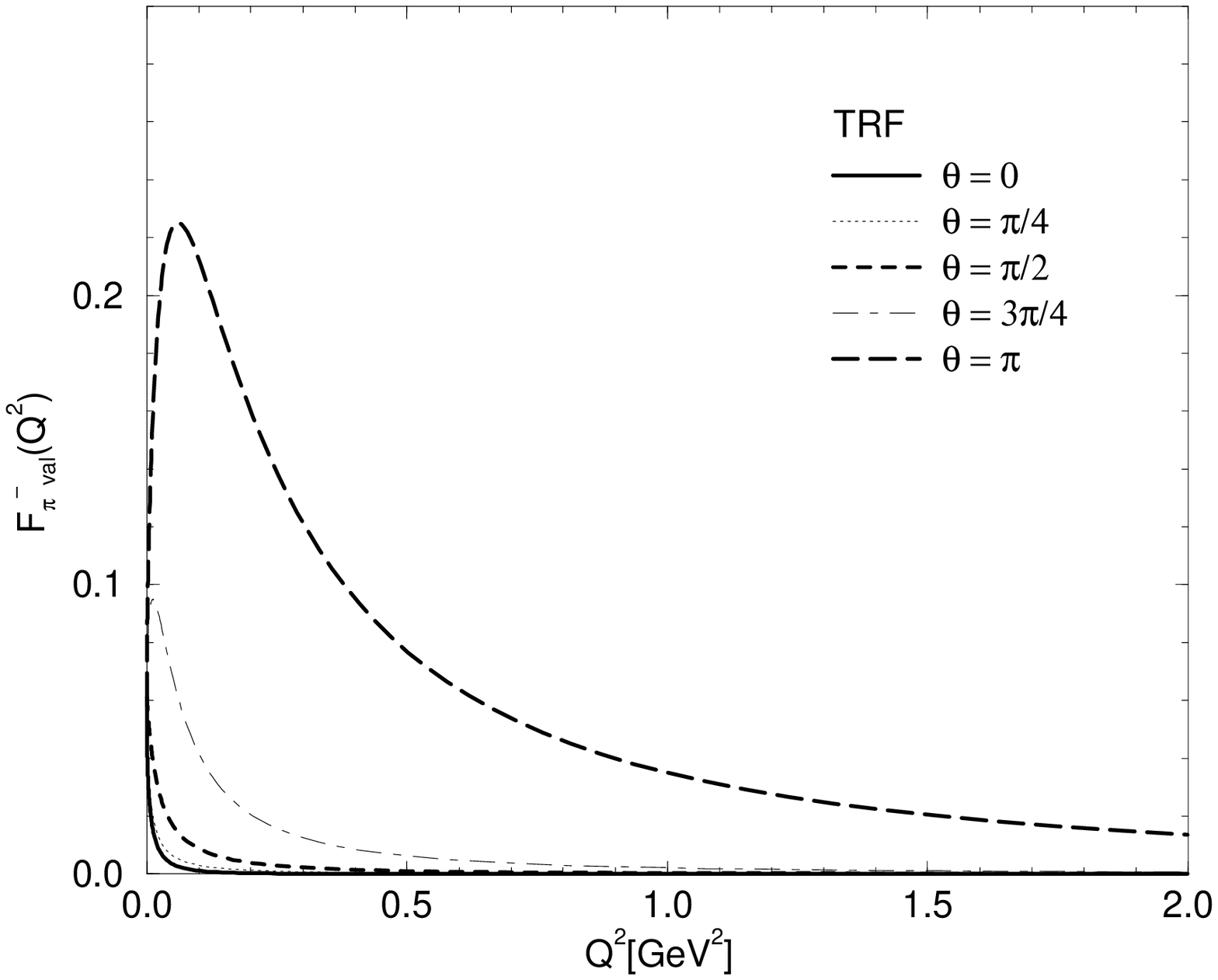,height=80mm,width=80mm}
 \epsfig{file=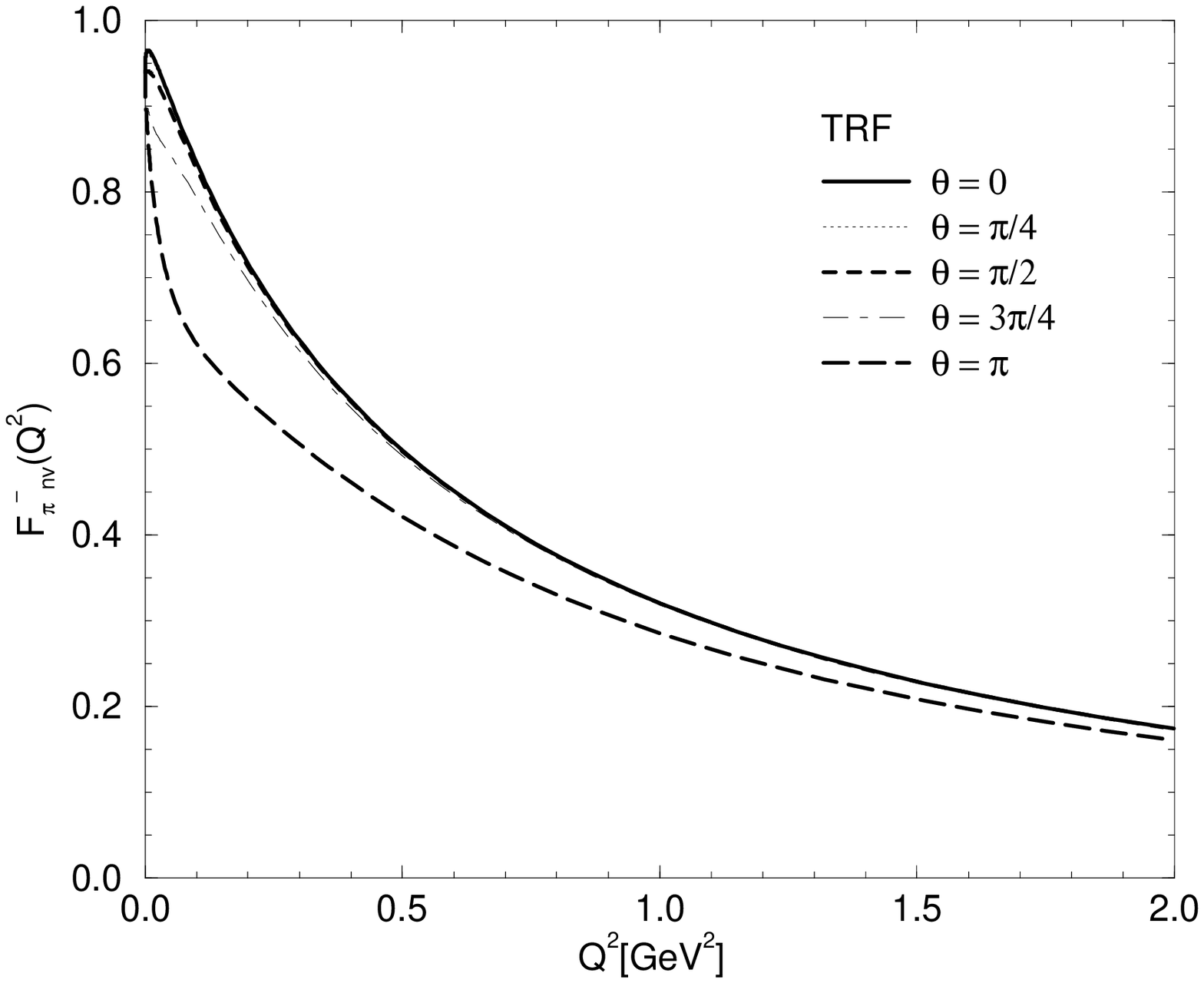,height=80mm,width=80mm}
\caption{Pion LF form factors $F^-_{\rm val}$ and $F^-_{\rm nv}$
 in the target-rest frame for five different
 values of the polar angle $\theta$. \label{fig0.7}}
\end{center}
\end{figure}
\begin{figure}
\begin{center}
 \epsfig{file=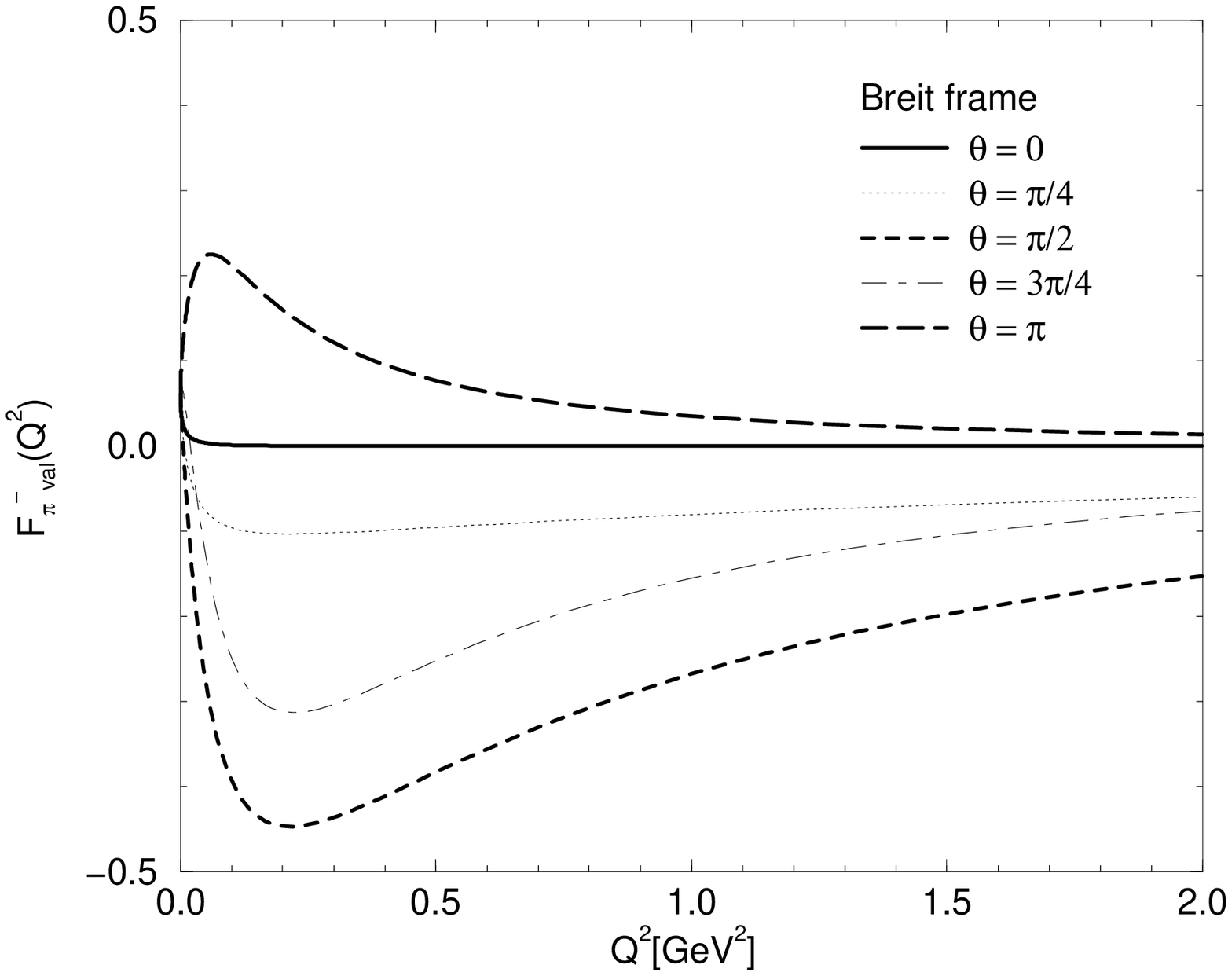,height=80mm,width=80mm}
 \epsfig{file=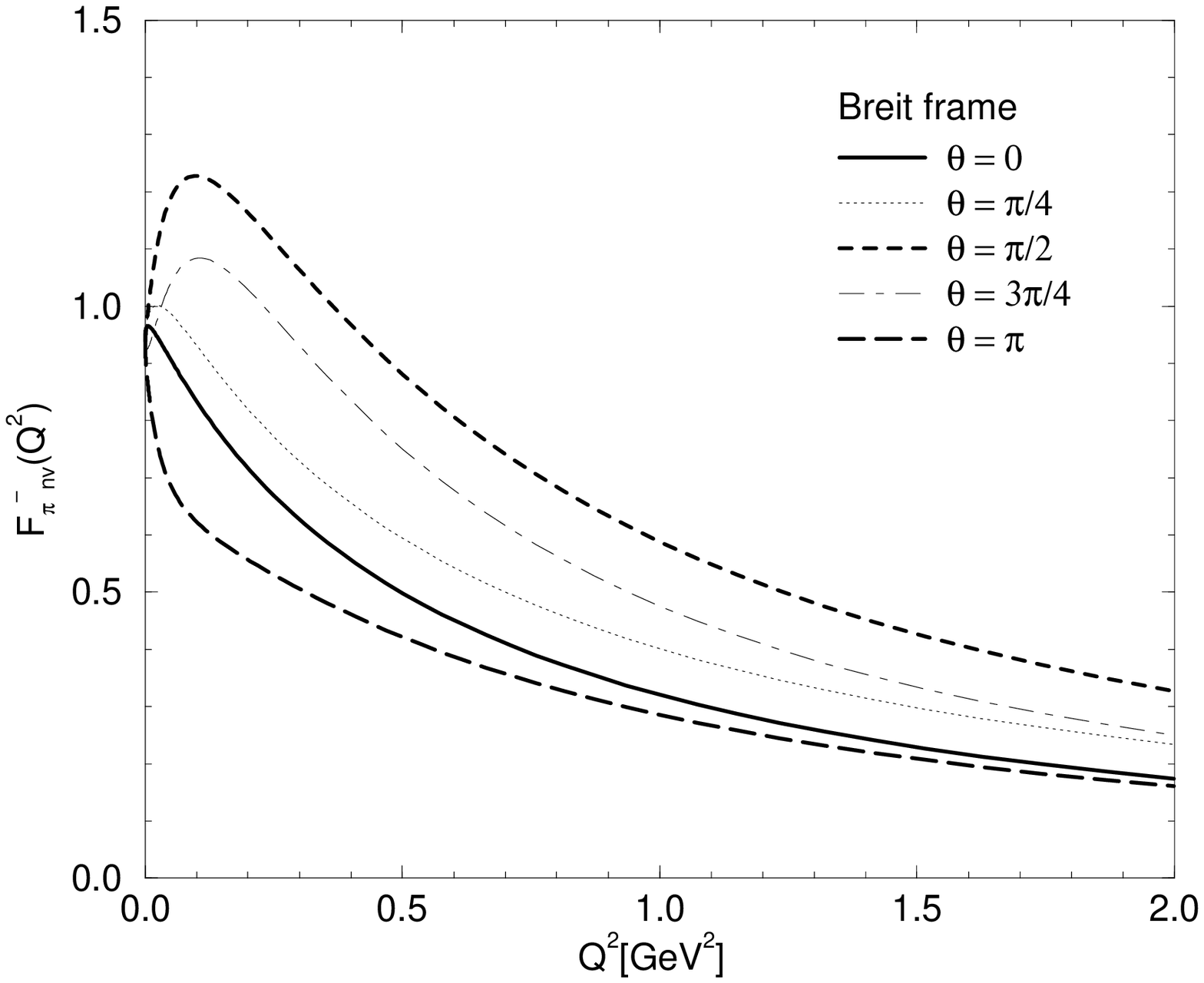,height=80mm,width=80mm}
\caption{Pion LF form factors $F^-_{\rm val}$ and $F^-_{\rm nv}$
 in the Breit frame for five different
 values of the polar angle $\theta$. \label{fig0.8}}
\end{center}
\end{figure}
\begin{figure}
\begin{center}
 \epsfig{file=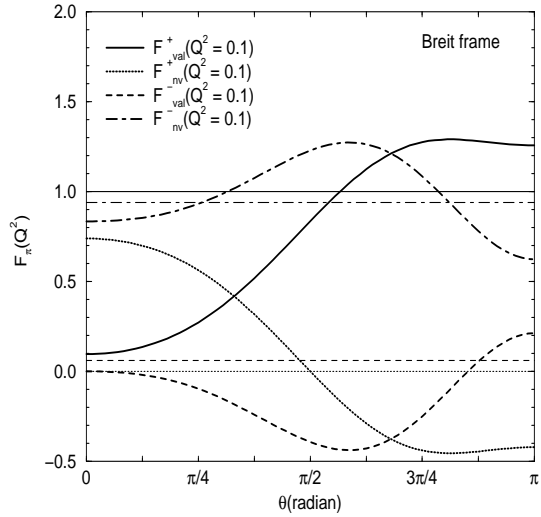,height=80mm,width=80mm}
\caption{Systematics of the angle dependence of the LF form factors
at $Q^2=0.1$ GeV$^2$(thick lines) and 
0 GeV$^2$(thin lines) in the Breit frame.\label{fig0.13}}
\end{center}
\end{figure}
\end{document}